\documentclass[fp,twocolumn]{jpsj3}
\usepackage{txfonts}

\usepackage{graphicx}
\usepackage{bm}
\usepackage{color}
\usepackage{mathrsfs}

\title{Possible High Thermoelectric Power in Semiconducting Carbon Nanotubes\\ 
{\it $\sim$ A Case Study of Doped One-Dimensional Semiconductors $\sim$}}

\author{Takahiro Yamamoto$^1$ and Hidetoshi Fukuyama$^2$}
\inst{$^1$ Department of Liberal Arts, Faculty of Engineering, Tokyo University of Science, Katsushika, Tokyo 125-8585, Japan \\
$^2$ Department of Applied Physics, Faculty of Science, Tokyo University of Science, Shinjuku, Tokyo 162-8601, Japan} 

\abst{
We have theoretically investigated the thermoelectric properties of impurity-doped one-dimensional semiconductors, focusing  
on nitrogen-substituted (N-substituted) carbon nanotubes (CNTs), using the Kubo formula combined with a self-consistent $t$-matrix 
approximation. N-substituted CNTs exhibit extremely high thermoelectric power factor ($PF$) values originating from a characteristic 
of one-dimensional materials where decrease in the carrier density increase both the electrical conductivity and the Seebeck coefficient 
in the low-N regime. The chemical potential dependence of the $PF$ values of semiconducting CNTs has also been studied as 
a field-effect transistor and it turns out that the $PF$ values show a noticeable maximum in the vicinity of the band edges. 
This result demonstrates that ``band-edge engineering" will be crucial for solid development of high-performance 
thermoelectric materials.
}
 
\kword{Seebeck coefficient, power factor, carbon nanotube, Kubo formula, self-consistent $t$-matrix approximation}

\begin{document}
\maketitle

\section{Introduction~\label{sec:1}}
Enhancing the performance of thermoelectric materials is an important issue to cope with future energy requirements. Hicks and 
Dresselhaus proposed in 1993 that significant enhancements could be realized by employing one dimensional (1D) 
thermoelectric materials~\cite{rf:hicks}. Subsequently, various nanowires and nanotubes exhibiting high thermoelectric 
performance were indeed discovered~\cite{rf:lin, rf:rabin, rf:heremans, rf:boukai, rf:hochbaum}. Carbon nanotubes (CNTs) are 
of particularly interest as high-performance, flexible and lightweight thermoelectric 1D materials
\cite{rf:small,rf:nakai,rf:hayashi1,rf:hayashi2,rf:Avery,rf:yanagi,rf:Shimizu,rf:nonoguchi1,rf:nonoguchi2,rf:nonoguchi3,rf:fujigaya1,rf:fujigaya2,rf:Jiang,rf:ohnishi}.

Recent experiments have shown that semiconducting CNTs exhibit sign inversion of the Seebeck coefficient from positive (p-type) to 
negative (n-type) upon the application of a gate voltage when using an electric double layer transistor in conjunction with an ionic 
liquid as the electrolyte~\cite{rf:yanagi,rf:Shimizu}. This result indicates the possibility of developing thermoelectric devices consisting 
solely of CNTs, since semiconducting CNTs exposed to the ambient atmosphere are usually p-type due to oxygen adsorption, and then
fabricating all-CNT thermoelectric devices will require synthesizing air-stable n-type CNTs.~\cite{rf:nonoguchi1,rf:nonoguchi2,rf:nonoguchi3,rf:fujigaya1,rf:fujigaya2}. 
However, the thermoelectric properties of impurity-doped n-type CNTs are not yet fully characterized. 

In the present study, we investigated the thermoelectric properties of nitrogen-doped (N-substituted) CNTs (a potential air-stable,
n-type CNT) on a theoretical basis, based on the linear response theory combined with a self-consistent $t$-matrix approximation to 
treat very disordered systems for which the Boltzmann transport theory is inadequate. These calculations indicate that N-substituted 
CNTs have an extremely high thermoelectric power factor resulting from a van Hove singularity at the conduction-band edge which is 
specific to 1D systems. Note that the results obtained in the present study can also be applied to boron-doped CNTs by replacing the 
impurity potential from an attractive potential to a repulsive one together with the change of conduction band to valence band.

\section{Theoretical Modeling and Formulation~\label{sec:2}}
\subsection{Linear response theory for thermoelectric effects~\label{sec:2.1}}
The thermoelectric effect is typically characterized by the Seebeck coefficient, $S$, which is defined as the voltage induced by 
a finite temperature gradient along a given direction (herein the $z$-direction) under the condition that there is no electrical current 
({\it i.e.}, $J=0$) along that direction. This can be written as
\begin{eqnarray}
S\equiv-\left(\frac{\Delta V}{\Delta T}\right)_{J=0},
\label{eq:S_def}
\end{eqnarray}
where $\Delta V$ is the induced voltage and $\Delta T$ is the temperature difference between the two ends of the material.

In the presence of both an electric field $\mathcal{E}$ and a temperature gradient $dT/dz$ along the $z$-direction, the current density 
$J$ is generally given by
\begin{eqnarray}
J=L_{11}\mathcal{E}-\frac{L_{12}}{T}\frac{dT}{dz}
\label{eq:J}
\end{eqnarray}
within the linear response with respect to $\mathcal{E}$ and $dT/dz$. The zero-current condition ($J=0$) leads to 
$L_{11}\mathcal{E}=\frac{L_{12}}{T}\frac{dT}{dz}$. Because the electric field and the temperature gradient can be written as 
$\mathcal{E}=-\Delta V/L$ and $dT/dz=\Delta T/L$ for a spatially uniform system with length $L$ (which we assume), 
$S$ as defined by Eq.~(\ref{eq:S_def}) can be expressed in terms of the response functions $L_{11}$ and $L_{12}$ as
\begin{eqnarray}
S=\frac{1}{T}\frac{L_{12}}{L_{11}}.
\label{eq:S}
\end{eqnarray}
It is important to note that $S$ depends on the electrical conductivity $L_{11}(=\sigma)$. 

One of the figures of merit for thermoelectric materials is the power factor $PF$, defined as
\begin{eqnarray}
PF\equiv\sigma S^2=\frac{1}{T^2}\frac{L_{12}^2}{L_{11}}.
\label{eq:PF}
\end{eqnarray}
Hence, the $PF$ values can be increased by increasing $L_{12}$ and decreasing $L_{11}$. 

At this point, it is helpful to discuss some of the history of the general and fully quantum expressions for $L_{11}$ and $L_{12}$. 
A general theory regarding the linear response to kinetic perturbation such as electric or magnetic fields was derived by Kubo in 
1956~\cite{rf:kubo}. However, a linear response theory for thermodynamic perturbations such as a temperature gradient was first 
proposed by Luttinger in 1964~\cite{rf:Luttinger} and has been futher developed since that time~\cite{rf:Smrcka,rf:Jonson_1980,rf:Kontani,rf:ogata-fukuyama}. 
In particular, in the case of non-interacting independent electrons scattered by static impurities without electron-phonon scattering, 
Jonson and Mahan~\cite{rf:Jonson_1980} have shown that $L_{11}$ and $L_{12}$ can be expressed by the following common 
function $\alpha(E)$, which is a zero-temperature conductivity represented by a current-current correlation function for a given energy $E$.
\begin{eqnarray}
L_{11}&=&\int_{-\infty}^\infty \!\!\!dE\left(-\frac{\partial f(E-\mu)}{\partial E}\right)\alpha(E),
\label{eq:L11}\\
L_{12}&=&-\frac{1}{e}\int_{-\infty}^\infty \!\!\!dE\left(-\frac{\partial f(E-\mu)}{\partial E}\right)(E-\mu)\alpha(E),
\label{eq:L12}
\end{eqnarray}
where $e$ is the elementary charge and $f(E-\mu)=1/(\exp((E-\mu)/k_{\rm B}T)+1)$ is the Fermi-Dirac distribution function. 
Hence, the Seebeck coefficient $S$ in Eq.~(\ref{eq:S}) and the power factor $PF$ in Eq.~(\ref{eq:PF}) can be determined from 
Eqs.~(\ref{eq:L11}) and (\ref{eq:L12}) once $\alpha(E)$ is known, although detailed studies are required to obtain this.

\subsection{Theoretical modeling of semiconducting CNTs~\label{sec:2.2}}
In this subsection, we briefly review the electronic structure of CNTs with zigzag-type chirality (z-CNTs). 
Within the nearest-neighbor-hopping $\pi$-orbital tight-binding approximation, the Hamiltonian of pristine z-CNTs 
without any defects and impurities is given by 
\begin{eqnarray}
\mathscr{H}_0=\sum_{q, k}\left(\epsilon_{qk}^{(+)}c_{qk}^{\dagger} c_{qk}
+\epsilon_{qk}^{(-)}v_{qk}^{\dagger} v_{qk}\right),
\label{eq:hamiltonian_0}
\end{eqnarray}
where $c_{qk}^\dagger$ and $v_{qk}^\dagger$ ($c_{qk}$ and $v_{qk}$) are the creation (annihilation) operators for 
the conduction- and valence-band electrons, respectively, $k$ is the wavenumber along the tube-axial direction, and 
$q$ is the discrete wavenumber along the circumferential direction. The energy dispersion $\epsilon_{qk}^{(\pm)}$ of 
the conduction $(+)$ and valence $(-)$ bands can be expressed as~\cite{rf:hamada, rf:saito}
\begin{eqnarray}
\epsilon_{qk}^{(\pm)}=\pm \gamma_0\sqrt{1+4\cos\left(\frac{ka_z}{2}\right)\cos\left(\frac{q\pi}{n}\right)+4\cos^2\left(\frac{q\pi}{n}\right)},\\
\left(q=0,1,\cdots, 2n-1 \quad{\rm and}~ -\pi/a_z<k<\pi/a_z\right)\nonumber.
\label{eq:TB_dispersion}
\end{eqnarray}
Here, $\gamma_0$ is the hopping integral between nearest-neighbor carbon atoms ($\pi$ orbitals, set to $\gamma_0=2.7eV$ 
in the present paper), $a_z=0.426$nm is the unit-cell length of z-CNTs, and $n$ is the natural number ($n=1, 2, \cdots, \infty$) 
specifying the unique structure of a particular z-CNT. Herein, the z-CNT with index $n$ is represented as ($n$, 0) CNT in 
accordance with customary practice. A ($n$, 0) CNT includes $4n$ carbon atoms in the unit cell and its diameter $d_{\rm t}$ is given by 
$d_{\rm t}=\frac{na_z}{\sqrt{3}\pi}$.

z-CNTs can be either metallic or semiconducting depending on whether or not $n$ is a multiple of 3, respectively.
In the case of metallic z-CNTs satisfying $n$~mod~$3=0$, two pairs of lowest-conduction (LC) and highest-valence (HV) bands 
$\epsilon_{qk}^{(\pm)}$ are specified by the following two values of $q$, respectively.
\begin{eqnarray}
q= \left\{
    \begin{array}{c}
      q_1\equiv 2n/3\\
      q_2\equiv 4n/3
    \end{array}
  \right. \quad {\rm for}\quad n~{\rm mod}~3=0.
\label{eq:pc0}
\end{eqnarray}
In contrast, in the case of semiconducting z-CNTs satisfying $n$~mod~$3\neq{0}$, the two pairs of LC and HV bands are respectively 
specified by 
\begin{eqnarray}
q= \left\{
    \begin{array}{c}
      q_1\equiv (2n+1)/3\\
      q_2\equiv (4n-1)/3
    \end{array}
  \right. \quad {\rm for}\quad n~{\rm mod}~3=1
\label{eq:pc1}
\end{eqnarray}
and
\begin{eqnarray}
q= \left\{
    \begin{array}{c}
      q_1\equiv (2n-1)/3\\
      q_2\equiv (4n+1)/3
    \end{array}
  \right. \quad {\rm for}\quad n~{\rm mod}~3=2.
\label{eq:pc2}
\end{eqnarray}
As seen in Eqs.~(\ref{eq:pc0})-(\ref{eq:pc2}), both the LC and HV bands will have two-fold degeneracy ($q_1$ and $q_2$) 
for a given $n$. 

The band gap $E_g$ of a z-CNT is equal to the energy difference $E_g=\epsilon_{q 0}^{(+)}-\epsilon_{q 0}^{(-)}$ ($q=q_1$, $q_2$)
between the LC and HV bands at $k=0$ and is calculated as
\begin{eqnarray}
E_g=2\gamma_0\left|1+2\cos\left(\frac{\pi q_i}{n}\right)\right| \quad (i=1, 2).
\label{eq:Eg}
\end{eqnarray}
In addition, the effective mass $m^*$ of an electron (a hole) in the two-fold degenerate LC (HV) bands is given by
\begin{eqnarray}
\frac{1}{m^*}&=&\frac{1}{\hbar^2}\left|\frac{\partial^2\epsilon_{qk}}{\partial k^2}\right|_{\substack{k=0,\\ q=q_i}}\\
&=&\frac{\gamma_0 a_z^2}{2\hbar^2}\left|\frac{\cos\left(\pi q_i/n\right)}{1+2\cos\left(\pi q_i/n\right)}\right| \quad (i=1, 2).
\label{eq:effectim_m}
\end{eqnarray}
From Eqs.~(\ref{eq:Eg}) and (\ref{eq:effectim_m}), we can readily determine that the band gap $E_{g}$ and the effective mass $m^*$ 
are zero for metallic z-CNTs satisfying $n$~mod~$3=0$, whereas these values will be finite for semiconducting z-CNTs satisfying 
$n$~mod~$3\neq{0}$. As an example, the $E_g$ and $m^*$ values of a semiconducting (10,0) CNT, which we focus on in the following, 
are estimated to be $E_g=0.95$eV and $m^*=0.093m_0$, where $m_0=9.11\times 10^{-31}$kg is the electron mass in a vacuum.

In n-type semiconducting z-CNTs, the transport phenomena are dominated by electrons in the vicinity of the LC-band bottom. 
In this case, the tight-binding Hamiltonian in Eq.~(\ref{eq:hamiltonian_0}) can be approximated by the effective-mass Hamiltonian
\begin{eqnarray}
\mathscr{H}^{\rm eff}_0=\sum_{q=q_1,q_2}\sum_{k}\epsilon_{k}c_{qk}^{\dagger} c_{qk},
\label{eq:H_eff}
\end{eqnarray}
where $\epsilon_{k}$ is the $q$-independent energy dispersion near the bottom of the LC band $\epsilon_{q k}^{(+)}$, 
which is given by
\begin{eqnarray}
\epsilon_{k}=\frac{\hbar^2k^2}{2m^*}.
\label{eq:k2-dispersion}
\end{eqnarray}
In Eq.~(\ref{eq:H_eff}), the energy origin ($E=0$~eV) is set at the bottom of the LC band. It should be noted that the 
effective-mass Hamiltonian in Eq.~(\ref{eq:H_eff}) is valid only for the case when the electrons are not thermally excited to the 
second-lowest conduction bands specified by $q=(2n-2)/3$ and $(4n+2)/3$ for $n$~{\rm mod}$~3=1$ and by
$q=(2n+2)/3$ and $(4n-2)/3$ for $n$~{\rm mod}$~3=2$. For a (10, 0) CNT, the energy difference $\Delta E$ between 
the bottom of the LC band and that of the second-lowest conduction band is $\Delta E=0.557$~eV. Herein, we will focus on 
the low-energy excitation regime in which the thermal energy $k_{\rm B}T$ is much lower than $\Delta E$.

At this point, we take account of a random potential term in $\mathscr{H}^{\rm eff}_0$ in Eq.~(\ref{eq:H_eff}) such that
\begin{eqnarray}
\mathscr{H}^{\rm eff}=\mathscr{H}^{\rm eff}_0+V_0\sum_{\langle i\rangle} c_i^\dagger c_i
\label{eq:H_eff+V0}
\end{eqnarray}
in order to examine the effects of N doping on z-CNTs. Here, $V_0$ is the attractive potential ($V_0<0$) of a N atom in
a z-CNT, {\it e.g.}, $V_0=-1.08$eV for a (10,0) CNT (see \S~\ref{sec:3.1} for details). In Eq.~(\ref{eq:H_eff+V0}), $c_{i}^\dagger$ ($c_{i}$) is 
the creation (annihilation) operator of an electron at the $i$th impurity site, and ${\langle i\rangle}$ represents the sum with respect to 
randomly distributed impurity positions with average concentration $c=N_{\rm imp}/N_{\rm unit}$, where $N_{\rm imp}$ is the total number of 
impurity sites and $N_{\rm unit}$ is the number of unit cells in a pristine CNT without any impurities. In the present work, we assume that 
the effects of possible mixing between the two LC bands due to impurity scattering can be ignored. In fact, this assumption is justified 
as will be shown in \S~\ref{sec:3.1}.

\subsection{Self-consistent $t$-matrix approximation~\label{sec:2.3}}
\begin{figure}[t]
  \begin{center}
  \includegraphics[keepaspectratio=true,width=75mm]{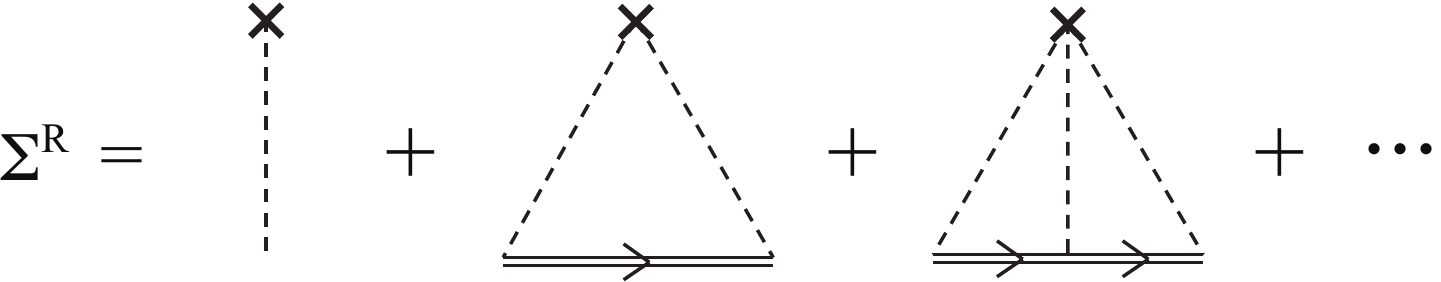}
  \end{center}
\caption{A self-consistent $t$-matrix approximation for the retarded self-energy of a one-particle retarded Green's function.
The $\times$-marks, the dotted lines and the solid double lines with arrow denote the impurity sites, the impurity potential and 
the one-particle retarded Green's function to be determined self-consistently, respectively.}
\label{fig:01}
\end{figure}

The modification of thermoelectric effects by randomly distributed impurities will be studied based on thermal Green's function 
formalism through the self-energy corrections of Green's functions and the coherent potential approximation (CPA), which is 
the most common practical approximation for this purpose~\cite{rf:velicky01,rf:velicky02}. Here, we employ a self-consistent 
$t$-matrix approximation~\cite{rf:klauder,rf:yonezawa,rf:hasegawa,rf:shiba}, which is a low-density limit ($c\to 0$) version of 
the CPA that has been shown to be applicable to low-doped semiconductors. This approximation proved to be also powerful 
in the recent study of spin-Seebeck effect~\cite{rf:ogata-fukuyama}. In this approximation, the retarded self-energy 
$\Sigma^{\rm R}(k,E)$ is determined by the processes shown in Fig.~\ref{fig:01}.
Here, we note that effects of Anderson localization~\cite{rf:nagaoka}, which play 
important roles at low temperature in one dimension but are not taken into account in this approximation, are negligible 
at high temperature of our main interest in this paper.

\begin{figure}[t]
  \begin{center}
  \includegraphics[keepaspectratio=true,width=70mm]{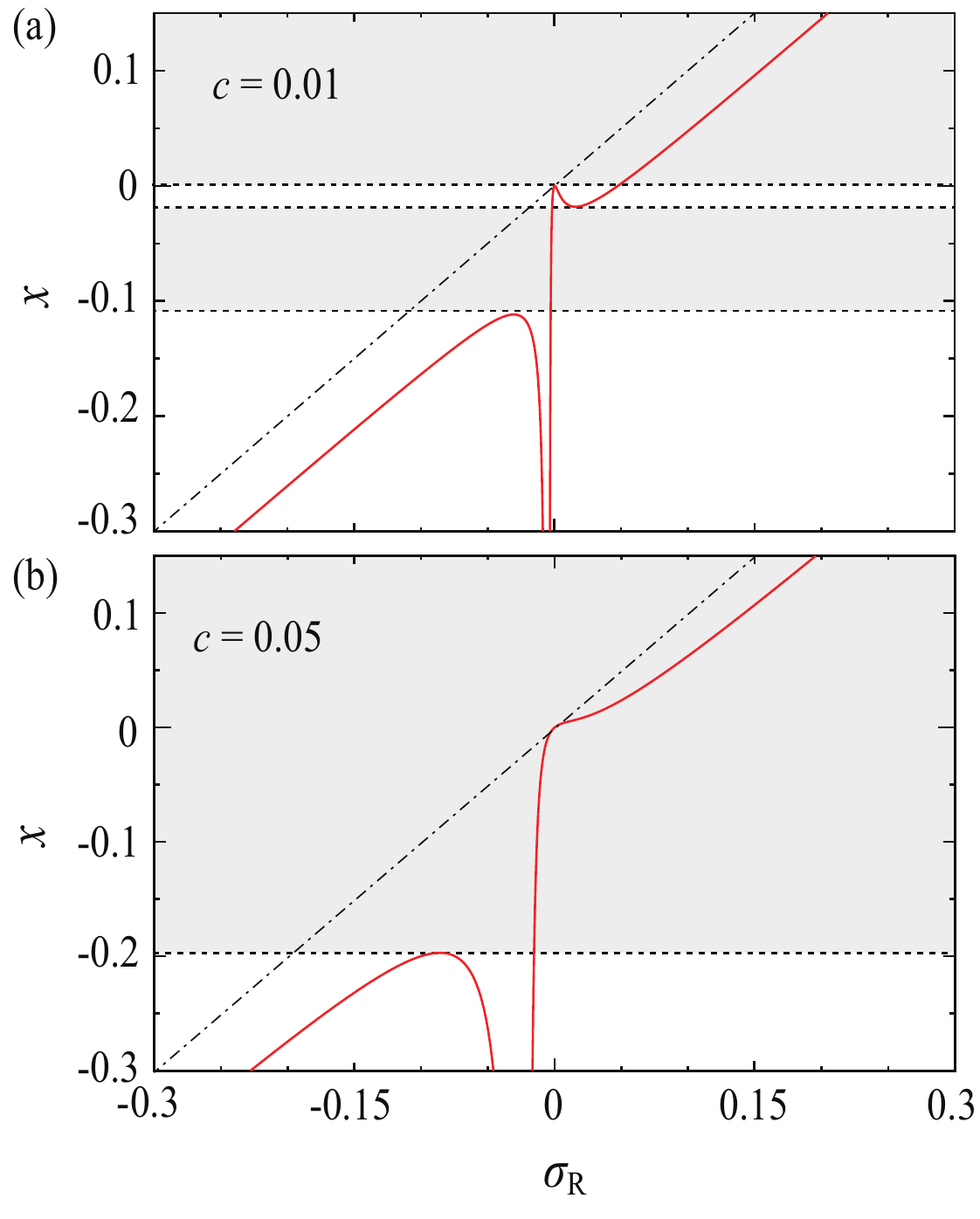}
  \end{center}
\caption{(Color online)  The $x$-$\sigma_{\rm R}$ relations for (a) $c=0.01$ and (b) $0.05$. The shaded regions indicate
the energy region where the cubic equation in Eq.~(\ref{eq:third_Sigma}) has one real solution and two complex solutions. 
The broken lines denotes $x=\sigma_{\rm R}$. }
\label{fig:02}
\end{figure}

Because of the short-range of the impurity potential in Eq.~(\ref{eq:H_eff+V0}), the retarded self-energy 
$\Sigma^{\rm R}(E)$ is independent of $k$ within the self-consistent $t$-matrix approximation and is determined by the requirement 
of self-consistency, as
\begin{eqnarray}
\Sigma^{\rm R}(E)=\frac{cV_0}{1-X(E)}, \quad {\rm Im}\Sigma^{\rm R}(E)<0
\label{eq:retarded_self-energy}
\end{eqnarray}
with 
\begin{eqnarray}
X(E)=\frac{V_0}{N_{\rm unit}}\sum_{k}\frac{1}{E-\epsilon_{k}-\Sigma^{\rm R}(E)}.
\label{eq:XE0}
\end{eqnarray}
By applying the effective-mass approximation in Eq.~(\ref{eq:k2-dispersion})
to Eq.~(\ref{eq:XE0}), the $k$-summation in Eq.~(\ref{eq:XE0}) can be analytically performed and we obtain
\begin{eqnarray}
X(x)=-\frac{i}{2}\frac{v_0}{\sqrt{x-\sigma_{\rm R}(x)}},\quad {\rm Im}\sqrt{x-\sigma_{\rm R}(x)}>0,
\label{eq:XE}
\end{eqnarray}
where $x\equiv E/t$, $v_0\equiv V_0/t$ and $\sigma_{\rm R}\equiv\Sigma^{\rm R}/t$ with
\begin{eqnarray}
t\equiv \frac{\hbar^2}{2m^*a_z^2}
\label{eq:t_ch}
\end{eqnarray}
equal to the characteristic energy of the z-CNT ({\it e.g.}, $t=2.26$eV for a (10,0) CNT). 
From Eqs.~(\ref{eq:retarded_self-energy}) and (\ref{eq:XE}), the self-consistent equation for $\sigma_{\rm R}(x)$ is given by
\begin{eqnarray}
\sigma_{\rm R}(x)=\frac{cv_0}{1+\frac{i}{2}\frac{v_0}{\sqrt{x-\sigma_{\rm R}(x)}}}.
\label{eq:self-consistent_Sigma}
\end{eqnarray}
Equation~(\ref{eq:self-consistent_Sigma}) can also be rewritten as 
\begin{eqnarray}
x=\sigma_{\rm R}-\frac{v_0^2\sigma_{\rm R}^2}{4(\sigma_{\rm R}-cv_0)^2}
\label{eq:x}
\end{eqnarray}
or as the cubic equation for $\sigma_{\rm R}$, 
\begin{eqnarray}
a_3\sigma_{\rm R}^3(x)+a_2\sigma_{\rm R}^2(x)+a_1\sigma_{\rm R}(x)+a_0=0
\label{eq:third_Sigma}
\end{eqnarray}
with $a_3=1$, $a_2=-(x+2cv_0+v_0^2/4)$, $a_1=cv_0(2x+cv_0)$, and $a_0=-x(cv_0)^2$.
Equation~(\ref{eq:third_Sigma}) indicates that for each energy, $x=E/t$, there are three solutions of $\sigma_{\rm R}(x)$:
three real ones or one real and two complex ones. These different cases are easily captured by showing Eq.~(\ref{eq:x})
as a function of real $\sigma_{\rm R}$,~\cite{rf:onodera,rf:fukuyama} as in Figs.~\ref{fig:02}(a) and \ref{fig:02}(b) for choices 
of $c=0.01$ and $c=0.05$, respectively. In the shaded regions of $x$, there are complex solutions of $\sigma_{\rm R}(x)$.

Once $\sigma_{\rm R}(E)$ is obtained via the above procedure, the density of states (DOS) can be determined
as follows. 
Within the self-consistent $t$-matrix and the effective-mass approximations, the DOS per the unit cell for each spin 
($\uparrow$ or $\downarrow$) and each orbital ($q_1$ or $q_2$) is given in terms of $\sigma_{\rm R}$ as
\begin{eqnarray}
\rho(x)&=&-\frac{1}{\pi V_0}{\rm Im}X(x)\label{eq:dos_dif}
\label{eq:DOS0}\\
&=&\frac{1}{2\pi t}{\rm Re}\frac{1}{\sqrt{x-\sigma_{\rm R}(x)}},~~ {\rm Im}\sqrt{x-\sigma_{\rm R}(x)}>0.
\label{eq:DOS}
\end{eqnarray}
Equation (\ref{eq:DOS}) indicates that for the region of $x$ with complex solutions of $\sigma_{\rm R}$(x), the DOS
will be finite, {\it i.e.}, in the shaded region in Figs.~\ref{fig:02}(a) and \ref{fig:02}(b). One caution is that the DOS
can be finite even for real $\sigma_{\rm R}$ if $x-\sigma_{\rm R}>0$; however all the solutions of $\sigma_{\rm R}(x)$
satisfying Eq.~(\ref{eq:x}) are in the region of $x-\sigma_{\rm R}(x)<0$ resulting in the absence of DOS (see also 
Figs.~\ref{fig:02}(a) and \ref{fig:02}(b)).

Similar to the DOS, the response functions $L_{11}$ and $L_{12}$ are also represented in terms of $\sigma_{\rm R}$
and eventually they can also be determined once $\sigma_{\rm R}(E)$ is obtained by solving Eq.~(\ref{eq:third_Sigma}).
The details of DOS, $L_{11}$, and $L_{12}$ of N-substituted z-CNTs will be discussed in the next section (\S~\ref{sec:3}).

\section{Numerical Results and Discussion~\label{sec:3}}

\subsection{Electronic states of N-substituted z-CNTs~\label{sec:3.1}}
Before discussing the effects of randomly-distributed N atoms in z-CNTs on their thermoelectric properties, we note that 
the impurity potential $V_0(<0)$ in the present 1D system results in a bound state for a single impurity. The binding energy 
$E=-E_{\rm b}(<0)$ of a single $N$ can be calculated from a pole of the $t$-matrix $\mathcal{T}(E)=V_0/(1-X(E))$, applying 
the limit of $c\to 0$, as
\begin{eqnarray}
E_{\rm b}=t \left(\frac{v_0}{2}\right)^2.
\end{eqnarray}
Because the binding energy of a single N atom in a (10,0) CNT is known to be $E_{\rm b}=(0.13\pm 0.02)$~eV based on 
first-principles calculations~\cite{rf:koretsune}, $V_0$ is set to $-1.08$eV for N-substituted (10,0) CNTs throughout this work.

At the last part of \S~\ref{sec:2.2}, we assumed that the effects of possible mixing between the two LC bands of z-SWNT can 
be ignored. This assumption is valid under the condition that the characteristic momentum $p_{\rm b}=\sqrt{2m^*E_{\rm b}}$ 
contributing to the formation of the bound state is much smaller than the momentum difference 
$\Delta p_{q}=\frac{2\pi\hbar}{w}|q_2-q_1|$ of the two LC bands, where $w=\sqrt{3}na_{\rm cc}$ is the circumference of 
a z-CNT and $a_{\rm cc}=0.142$nm is the C-C bond length.

\begin{figure}[t]
  \begin{center}
  \includegraphics[keepaspectratio=true,width=70mm]{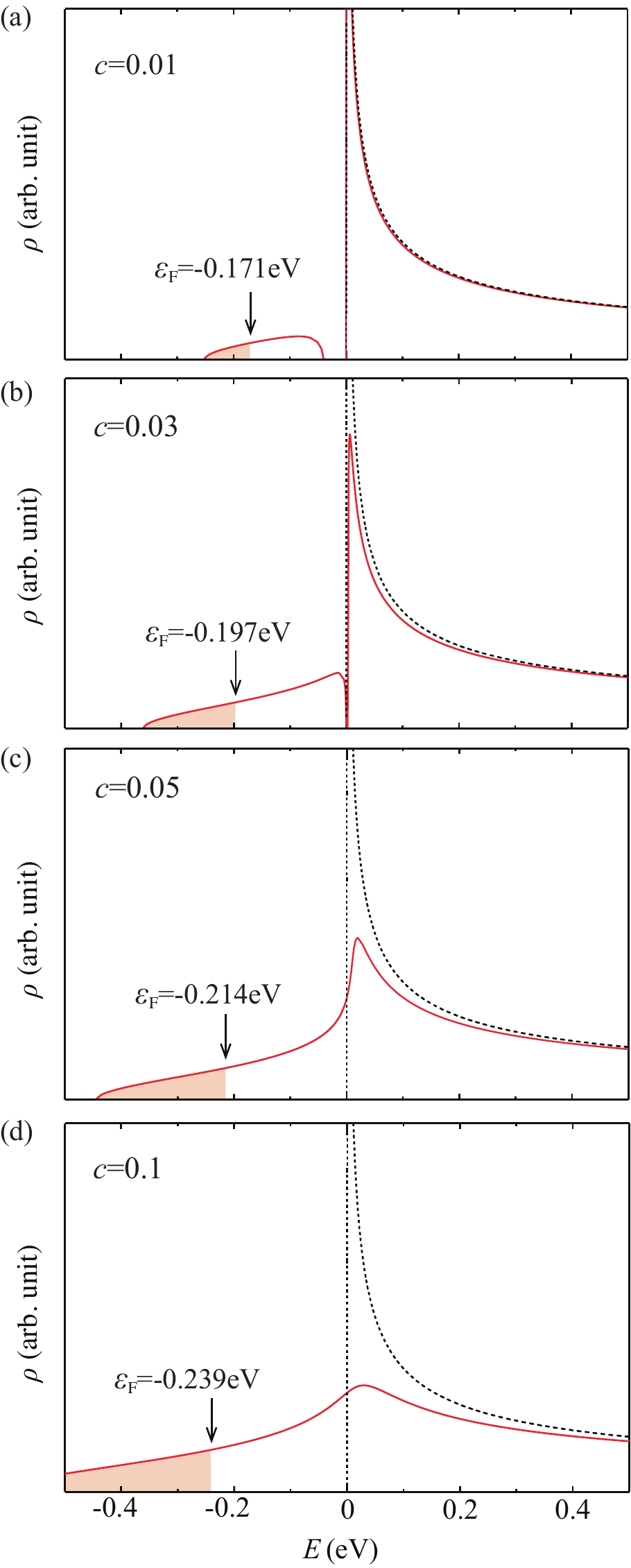}
  \end{center}
\caption{(Color online) Density of states (DOS) results for a (10,0) CNT for which $c=0.01$ (a), $0.03$ (b), $0.05$ (c), and 
$0.1$ (d). The solid and dotted curves show the DOS results for N-substituted and pristine (10,0) CNTs. The energy origin ($E=0$eV) 
was set equal to the bottom of conduction band of pristine (10,0) CNTs. The arrows indicate the Fermi energy.}
\label{fig:03}
\end{figure}

Figure~\ref{fig:03} presents the DOS results for (10,0) CNTs with various concentrations of N atoms ($c=0.01$, $0.03$, $0.05$, and $0.1$). 
Each arrow in Figs.~\ref{fig:03}(a)-\ref{fig:03}(d) indicates the Fermi energy $\epsilon_{\rm F}$. With increasing $c$, $\epsilon_{\rm F}$ 
shifts downward from $E=-E_{\rm b}$. As shown by the solid curves in Figs.~\ref{fig:03}(a) and \ref{fig:03}(b), the impurity band centered 
around $E=-E_{\rm b}(=-0.13~{\rm eV})$ is separated from the conduction band (referred to as {\it a persistence type DOS}~\cite{rf:onodera}). 
In constast, those in Figs.~\ref{fig:03}(c) and \ref{fig:03}(d) indicate that the impurity band merges into the conduction band 
({\it an amalgamation type DOS}~\cite{rf:onodera}). Boundary between the persistence and amalgamation types is determined by 
the critical value of $c$, which is given by
\begin{eqnarray}
c_{\rm critical}=\frac{2}{27}|v_0|
=\frac{4}{27}\sqrt{\frac{E_{\rm b}}{t}}
\label{eq:c_critical}
\end{eqnarray}
In the case of N-substituted (10,0) CNT, $c_{\rm critical}$ is estimated to be $c_{\rm critical}=0.035$. 
The derivation of Eq.~(\ref{eq:c_critical}) is explained later in this subsection.

The dotted curves in Fig.~\ref{fig:03} represent the DOS results for pristine (10,0) CNTs without any defects or impurities,
and these exhibit a van Hove singularity at the conduction band edge at $E=0$~eV (see also Eq.~(\ref{eq:rho_0E_zCNT}) 
in \S~\ref{sec:3.5}). Although the van Hove singularity disappears in the presence of N impurities, the persistent-type DOS results 
in Figs.~\ref{fig:03}(a) and \ref{fig:03}(b) exhibit a sharp peak near $E=0$~eV, implying that the electrons in the conduction bands 
are not greatly scattered by the N impurities in these cases. Conversely, the amalgamation-type DOS results in Figs.~\ref{fig:03}(c) 
and \ref{fig:03}(d) are considerably spread out in the vicinity of $E=0$~eV. This occurs because conduction-electron and impurity 
states are strongly mixed near the conduction band edge ($E=0$~eV) in the case of an amalgamation-type DOS with $c>c_{\rm critical}$.

\begin{figure}[t]
  \begin{center}
  \includegraphics[keepaspectratio=true,width=75mm]{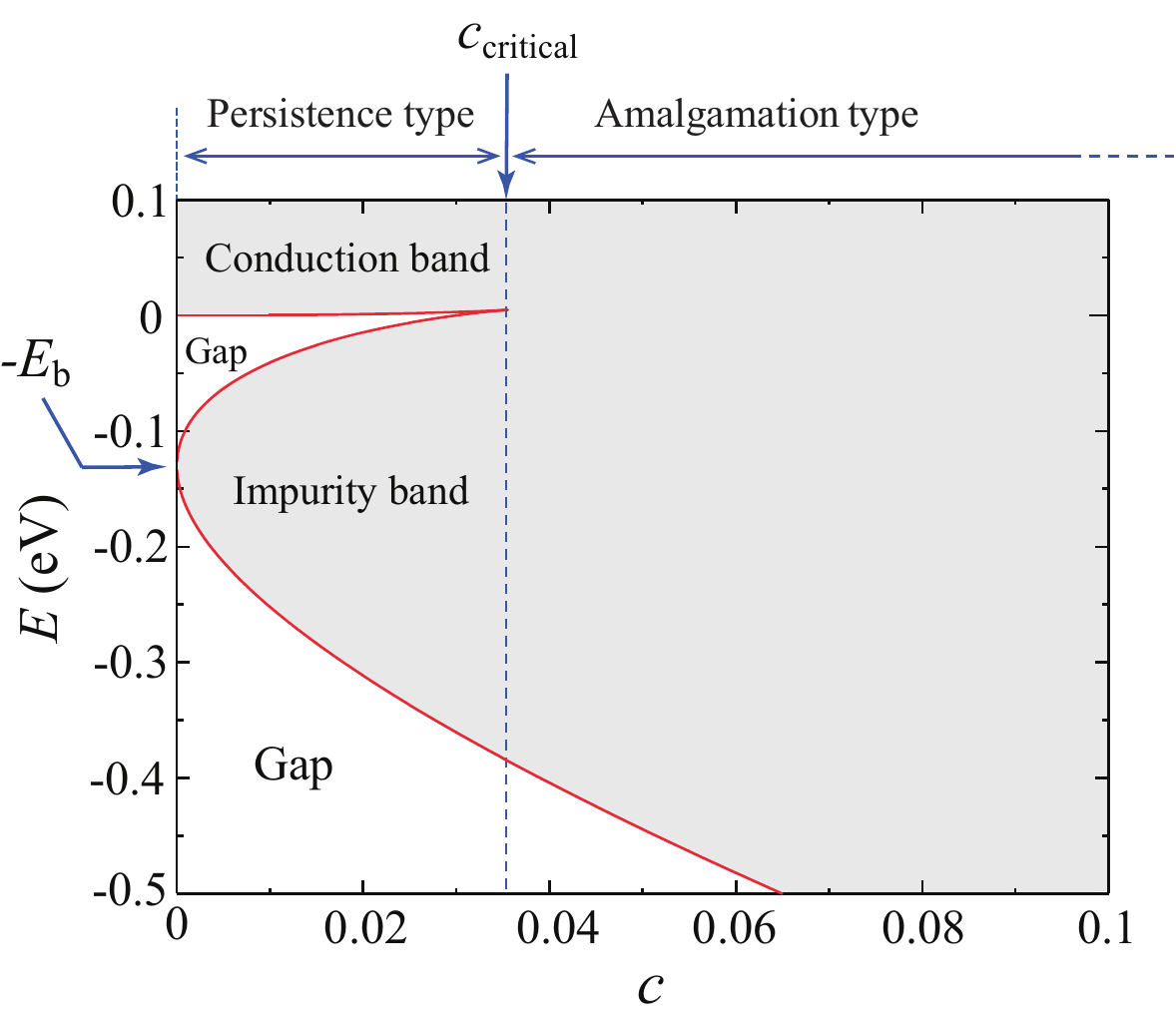}
  \end{center}
\caption{(Color online) A DOS diagram on the $c$-$E$ plane showing the presence or absence of a DOS for N-substituted 
(10,0) CNTs. The shaded region indicates the region where the finite DOS exists, whereas the unshaded regions are
the gap region where the DOS is zero.}
\label{fig:04}
\end{figure}

Here, we show a DOS diagram on the $c$-$E$ plane indicating the presence or absence of a finite DOS for N-substituted 
(10,0) CNTs in Fig.~\ref{fig:04}. The boundary between the finite- and zero-DOS regions in Fig.~\ref{fig:04} can be determined 
from the condition $dx/d\sigma_{\rm R}=0$ (see Fig.~\ref{fig:02}). In fact, by imposing the condition $dx/d\sigma_{\rm R}=0$ to 
Eq.~(\ref{eq:x}), we obtain the cubic equation
\begin{eqnarray}
b_3\sigma_{\rm R}^3+b_2\sigma_{\rm R}^2+b_1\sigma_{\rm R}+b_0=0
\label{eq:b}
\end{eqnarray}
with $b_3=1$, $b_2=-3cv_0$, $b_1=cv_0^2(3c+v_0/2)$, and $b_0=-c^3v_0^3$. Substituting the real solutions of 
Eq.~(\ref{eq:b}) to Eq.~(\ref{eq:x}), the boundary indicated by the solid curve in Fig.~\ref{fig:04} can be obtained. 
In addition, $c_{\rm critical}$ in Eq.~(\ref{eq:c_critical}) can be derived from the condition that the discriminant 
$D=-(27/4)c^3v_0^8(c-2|v_0|/27)$ for Eq.~(\ref{eq:b}) is equal to zero ($D=0$).

\begin{figure}[t]
  \begin{center}
  \includegraphics[keepaspectratio=true,width=70mm]{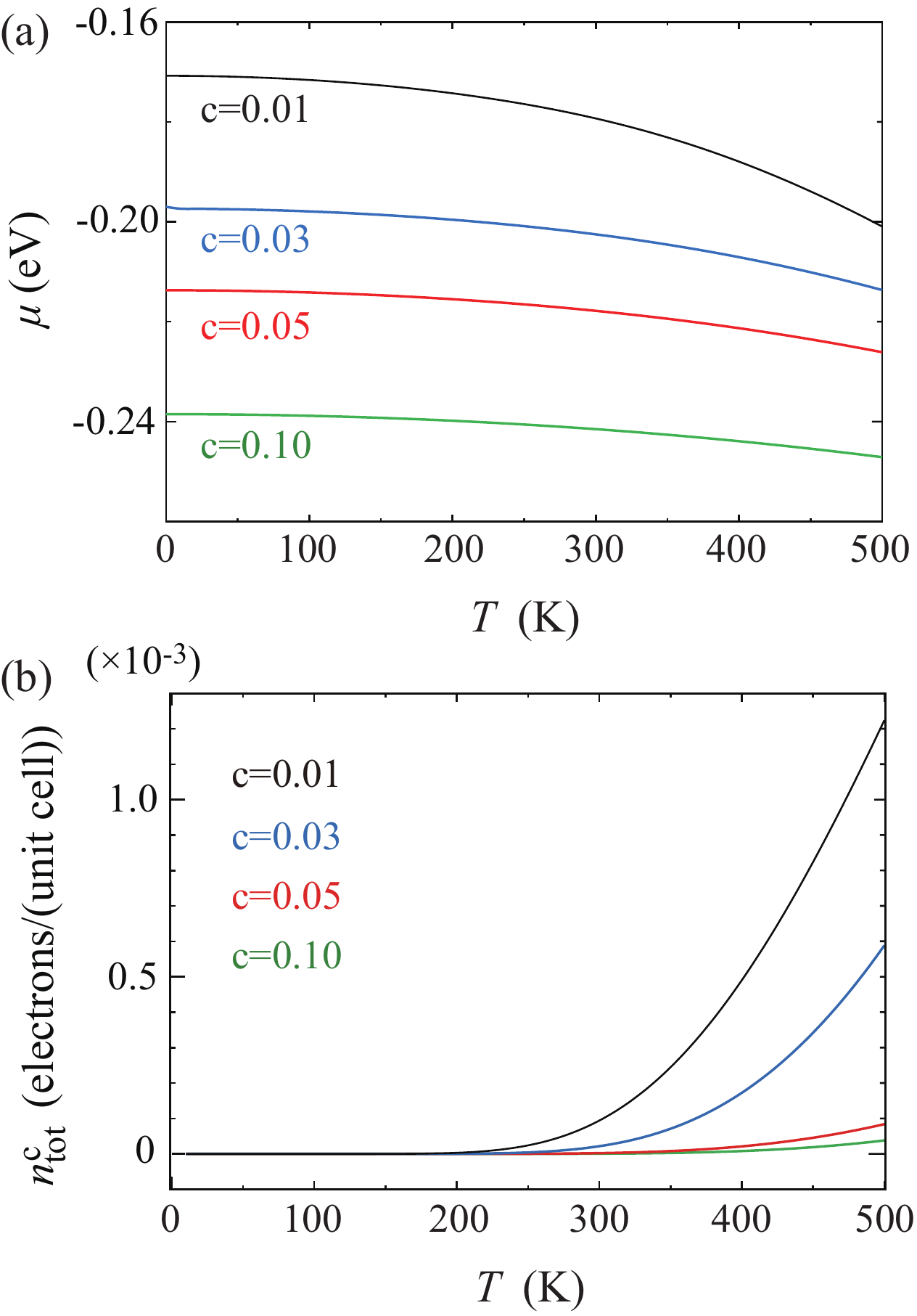}
  \end{center}
\caption{(Color online) The temperature dependence of (a) the chemical potentials $\mu$ and (b) the electron density 
$n^{\rm c}_{\rm tot}$ in the conduction-band region ($E\ge 0$~eV) of N-substituted (10,0) CNTs for which $c=0.01$, $0.03$, $0.05$, 
and $0.1$.}
\label{fig:05}
\end{figure}

Once the density of states, $\rho(x)$, in Eq.~(\ref{eq:DOS}) is obtained, the chemical potential $\mu$ is determined from 
the relationship between $\rho(E)$ and the electron density $n$ per unit cell for each spin ($\uparrow$ or $\downarrow$) 
and each orbital ($q_1$ or $q_2$) as
\begin{eqnarray}
n=\int_{-\infty}^{\infty}\!\!\! dE\rho(E) f(E-\mu).
\label{eq:DOS2}
\end{eqnarray}
For N-substituted z-CNTs, the total electron density $n_{\rm tot}=4n$ is set equal to the impurity concentration $c$ ({\it i.e.}, $n=c/4$) 
because each N substituting a C-site supplies a single electron to the CNT. It is noted that the factor 4(=2$\times$2) originates from 
the spin and orbital degeneracies. Figure~\ref{fig:05}(a) presents the temperature dependence of $\mu$ for a N-substituted (10,0) 
CNT for which $c=0.01$, $0.03$, $0.05$ and $0.1$. It is seen that $\mu$ decreases monotonically from $\mu=\epsilon_{\rm F}$ as 
the temperature increases. In these cases of $0.01\le c\le 0.1$, $\mu$ lies in the impurity band even at $T=500$~K and the systems 
are in extrinsic regime where the N atoms are partially ionized (see also Fig.~\ref{fig:08}). We wish to emphasize that conventional 
Boltzmann transport theory is totally inadequate for this impurity-band conduction, and that the Kubo formula together with the CPA 
has a significant merit in such a situation. Figure~\ref{fig:05}(b) shows the temperature dependence of the electron density 
$n^{\rm c}_{\rm tot}$ in the conduction-band region
$E\ge 0$~eV, which is given by $n^{\rm c}_{\rm tot}=4n_{\rm c}$ with
\begin{eqnarray}
n_{\rm c}=\int_{0}^{\infty}\!\!\! dE\rho(E) f(E-\mu).
\label{eq:DOS3}
\end{eqnarray}
As seen in Fig.~\ref{fig:05}(b), the characteristic temperature at which $n_{\rm c}$ begins to rise up becomes lower with decreasing $c$,
because $\epsilon_F$ at $T=0$~K locates closer to the band edge ($E=0$~eV) as $c$ get smaller as seen from Fig.~\ref{fig:03}.

\subsection{Temperature dependence of $L_{11}$ and $L_{12}$~\label{sec:3.2}}

In this subsection, we discuss the temperature dependence of the $L_{11}$ and $L_{12}$ values of N-substituted (10,0) CNTs. 
Within the self-consistent $t$-matrix approximation, $\alpha(E)$ can be expressed as~\cite{rf:Jonson_1980,rf:ogata-fukuyama} 
\begin{eqnarray}
\alpha(E)=4\frac{e^2\hbar}{\pi V}\sum_{k} v_{k}^2\left[{\rm Im~} G^{\rm R}(k, E)\right]^2,
\label{eq:alpha_E}
\end{eqnarray}
where the factor 4 comes from the spin and orbital degeneracies,
and $V$ is a volume of a system. Here, $G^{\rm R}(k, E)$ is the retarded Green's function
\begin{eqnarray}
G^{\rm R}(k, E)=\frac{1}{E-\epsilon_{k}-\Sigma^{\rm R}(E)}.
\label{eq:retarded_G}
\end{eqnarray}
Furthermore, within the effective-mass approximation for z-CNTs in Eq.~(\ref{eq:k2-dispersion}), the $k$-summation in Eq.~(\ref{eq:alpha_E}) 
can be performed analytically and $\alpha(x)$ is given by
\begin{eqnarray}
\alpha(x)=\frac{2e^2}{\pi\hbar}\frac{a_z}{A}\frac{\left({\rm Re}\sqrt{x-\sigma_{\rm R}(x)}\right)^2}{|x-\sigma_{\rm R}(x)|{\rm Im}\sqrt{x-\sigma_{\rm R}(x)}},
\label{eq:alpha_eff-m}
\end{eqnarray}
where $A$ is the cross-sectional area of a z-CNT ($A\equiv\pi d_{\rm t}\delta$ is conventionally used as the effective cross-sectional area 
of a CNT, where $\delta=0.34$nm is the van der Waals diameter of carbon). Substituting Eq.~(\ref{eq:alpha_eff-m}) into Eqs.~(\ref{eq:L11}) 
and (\ref{eq:L12}), $L_{11}$ and $L_{12}$ can be calculated.

\begin{figure}[t]
  \begin{center}
  \includegraphics[keepaspectratio=true,width=70mm]{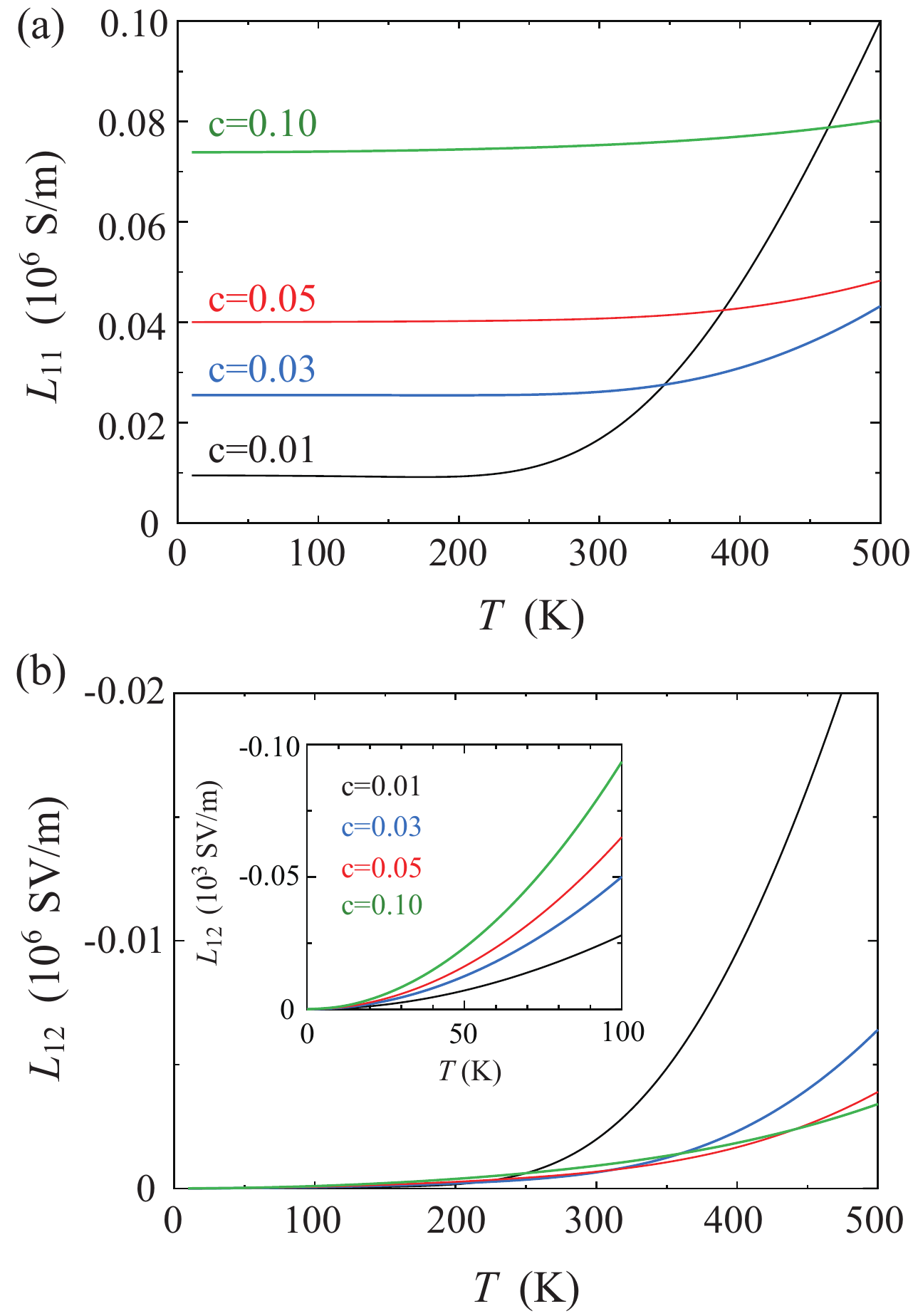}
  \end{center}
\caption{(Color online) The temperature dependence of (a) electrical conductivity $L_{11}$ and (b) the thermoelectric coefficient 
$L_{12}$ of N-substituted (10,0) CNTs for which $c=0.01$, $0.03$, $0.05$, and $0.1$. Note that the vertical axis of Fig.~\ref{fig:04}(b) 
is negative.}
\label{fig:06}
\end{figure}

Figure~\ref{fig:06}(a) shows the $T$ dependence of the $L_{11}$ value of N-substituted (10,0) CNTs for which $c=0.01$, $0.03$, $0.05$, 
and $0.1$. In the low-$T$ region, $L_{11}$ is almost constant with respect to temperature. With increasing $T$, $L_{11}$ increases 
above this constant value. As seen from the comparison between Fig.~\ref{fig:05}(b) and Fig.~\ref{fig:06}(a), $L_{11}$ begins to 
increase rapidly when $n_{\rm c}$ rises. The rapid increase of $L_{11}$ results from the long life-time of 
conduction-band electrons in the vicinity of the LC band bottom (see also Fig.~\ref{fig:09} in \S~\ref{sec:3.5}). This behavior of 
$L_{11}$ can be quantitatively understood using Eq.~(\ref{eq:L11}). Specifically, the Sommerfeld expansion of Eq.~(\ref{eq:L11}) allows 
the low-temperature $L_{11}$ behavior to be expressed as
\begin{eqnarray}
L_{11}\approx
\alpha(\epsilon_{\rm F})+\frac{(\pi k_{\rm B}T)^2}{6}\alpha''(\epsilon_{\rm F}),
\label{eq:L11_low-T}
\end{eqnarray}
We can see from Eq.~(\ref{eq:L11_low-T}) that $L_{11}$ deviates from the constant value $\alpha(\epsilon_{\rm F})$ in proportion to $T^2$.
It should be noted that the temperature region where Eq.~(\ref{eq:L11_low-T}) is applicable is some fraction of the width of the impurity band.
In other words, the rapid increases of $L_{11}$ for $c=0.01$ above $\sim200$K as shown in Fig.~\ref{fig:06}(a) cannot be described by
Eq.~(\ref{eq:L11_low-T}).

Figure~\ref{fig:06}(b) summarizes the $T$ dependence of $L_{12}$ for N-substituted (10,0) CNTs for which $c=0.01$, $0.03$, $0.05$, 
and $0.1$. $L_{12}$ increases monotonically from zero with increasing $T$. Similar to the above discussion regarding $L_{11}$, 
the low-$T$ behavior of $L_{12}$ can be quantitatively understood by performing the Sommerfeld expansion of Eq.~(\ref{eq:L12}) 
as follows. At low $T$, $L_{12}$ can be expressed as
\begin{eqnarray}
L_{12}\approx-\frac{(\pi k_{\rm B}T)^2}{3e}\alpha'(\epsilon_{\rm F})
\label{eq:L12_low-T}
\end{eqnarray}
up to the lowest order of $T$ in the Sommerfeld expansion of Eq.~(\ref{eq:L12}). In fact, the $T^2$ behavior of $L_{12}$ at low $T$ 
can be seen in the inset to Fig.~\ref{fig:06}(b). In addition, we see that $L_{12}$ at low $T$ increases with respect to $c$ because
$\alpha'(\epsilon_{\rm F})$ in Eq.~(\ref{eq:L12_low-T}) is a monotonically increasing function of $c$ as shown in Fig.~\ref{fig:A1}(b) 
in Appendix. In contrast, $L_{12}$ at high $T$ decrease with respect to c, since impurity scattering of electrons thermally excited to 
conduction bands increases with respect to $c$.

\subsection{Temperature dependence of $S$ and $PF$~\label{sec:3.3}}

\begin{figure}[t]
  \begin{center}
  \includegraphics[keepaspectratio=true,width=70mm]{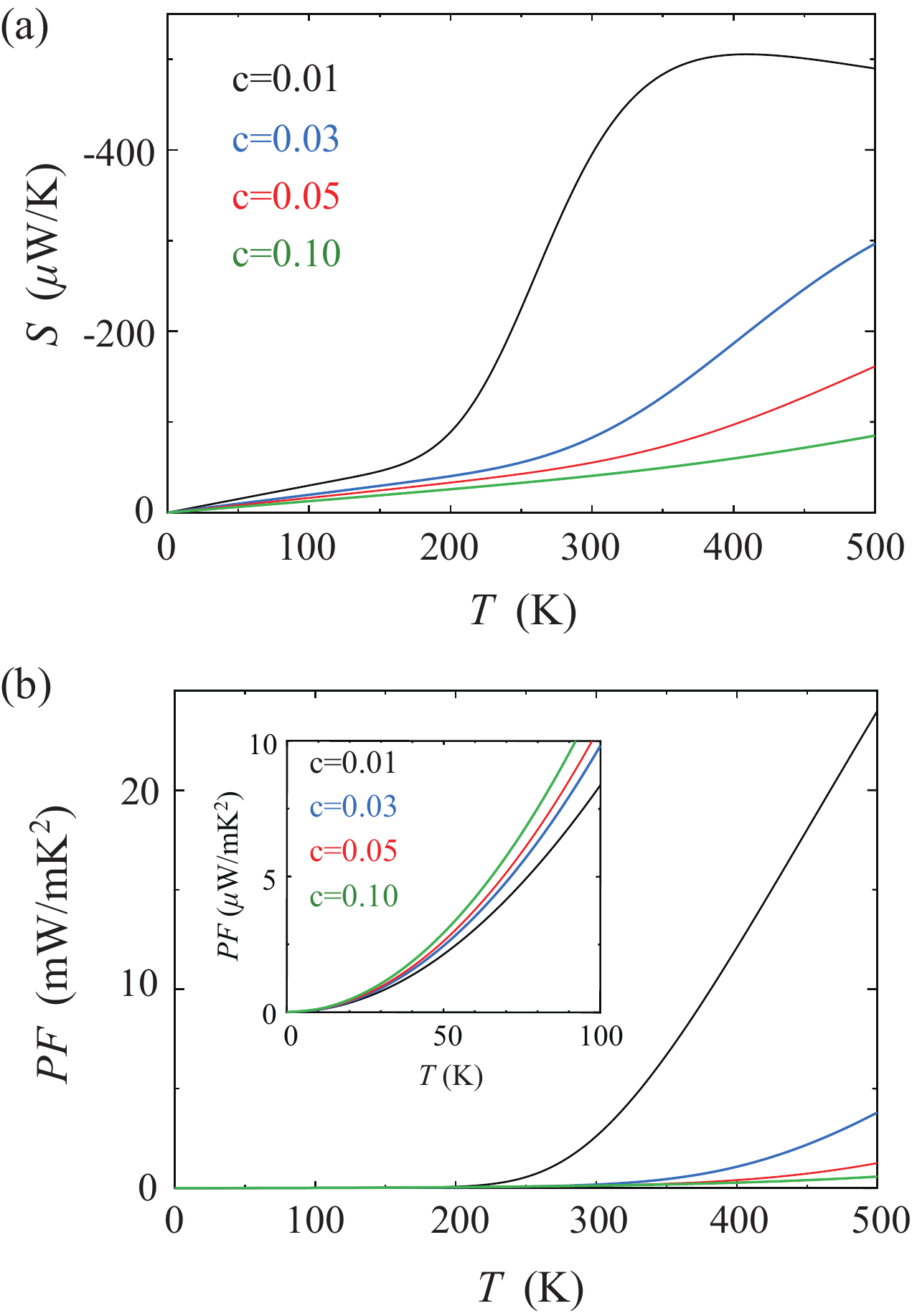}
  \end{center}
\caption{(Color online) The temperature dependence of (a) the Seebeck coefficient $S$ and (b) the power factor $PF$ of 
N-substituted (10,0) CNT for which $c=0.01$, $0.03$, $0.05$, and $0.1$. Note that the vertical axis of Fig.~\ref{fig:06}(a) is negative.}
\label{fig:07}
\end{figure}

The Seebeck coefficient $S$ can be calculated using Eq.~(\ref{eq:S}), and Fig.~\ref{fig:07}(a) plots the $T$ dependence of 
the $S$ values of N-substituted (10,0) CNTs for which $c=0.01$, $0.03$, $0.05$, and $0.1$. In the low-$T$ region, 
substituting Eqs.~(\ref{eq:L11_low-T}) and (\ref{eq:L12_low-T}) into Eq.~(\ref{eq:S}) and expanding the equation up to the lowest order 
of $T$, we see
\begin{eqnarray}
S=-\frac{\pi^2 k_{\rm B}}{3e}\frac{\alpha'(\epsilon_{\rm F})}{\alpha(\epsilon_{\rm F})}k_{\rm B}T.
\label{eq:S_low-T}
\end{eqnarray}
This is formally the same as the Mott formula, except that here $\alpha(\epsilon_{\rm F})$ represents the conductivity in the impurity 
band in contrast to the conductivity of band electrons treated by the Boltzmann transport equation in the original Mott formula~\cite{rf:mott}. 
The general validity of the Mott formula has already been demonstrated by Jonson and Mahan on the basis of linear response theory 
in terms of the thermal Green's function~\cite{rf:Jonson_1980}. The present results are based on a particular model and approximation
and are in complete accordance with these previous works. One note of caution is that, in the present study, the temperature region 
over which the formula in Eq.~(\ref{eq:S_low-T}) is valid will be some fraction of the width of the impurity band, in contrast to the Fermi 
energy in the original Mott formula. In fact, $S$ is proportional to $T$ only when $n^{\rm c}_{\rm tot}\approx 0$, as seen from comparison 
between Fig.~\ref{fig:05}(b) and Fig.~\ref{fig:07}(a). 
On the other hand, the large $|S|$ is realized in higher temperature region as seen in Fig.~\ref{fig:07}(a) where the Mott formula fails.

Similarly, the power factor $PF$ can be calculated using Eq.~(\ref{eq:PF}). Figure~\ref{fig:07}(b) shows the temperature dependence of 
the $PF$ values of N-substituted (10,0) CNTs for which $c=0.01$, $0.03$, $0.05$, and $0.1$. In the inset to Fig.~\ref{fig:07}(b), 
the low-temperature $PF$ is proportional to $T^2$ as 
\begin{eqnarray}
PF=\left(\frac{\pi^2 k_{\rm B}}{3e}\right)^2\frac{\alpha'^2(\epsilon_{\rm F})}{\alpha(\epsilon_{\rm F})}(k_{\rm B}T)^2,
\label{eq:PF_low-T}
\end{eqnarray}
which is obtained by substituting Eqs.~(\ref{eq:L11_low-T}) and (\ref{eq:L12_low-T}) into Eq.~(\ref{eq:PF}).

We note from Fig.~\ref{fig:07}(b) that the $PF$ at high $T$ becomes large with decreasing $c$, a trend that is the exactly opposite to 
the $c$ dependence of $PF$ at low $T$. As an example, the $PF$ of N-substituted (10,0) CNTs for which $c=0.01$ has a large value 
on the order of 1~mW/mK$^2$ at room $T$. In the next subsection (\S~\ref{sec:3.4}), we discuss the $PF$ values of N-substituted 
(10,0) CNTs with smaller $c$ below $c<0.01$ in detail.

\subsection{N-concentration dependence of $\sigma$, $S$ and $PF$~\label{sec:3.4}}
In this subsection, we discuss the $c$-dependence of $\sigma$, $S$ and $PF$ at $T=300$K, $400$K and $500$K,
which are calculated based on the self-consistent $t$-matrix approximation.
Because the maximum N concentration that can be doped into a CNT is limited to approximately $1$\%~\cite{ref:Glerup} at present,
we focus here on the $c\le 0.01$ values representing low N concentration.

\begin{figure}[t]
  \begin{center}
  \includegraphics[keepaspectratio=true,width=70mm]{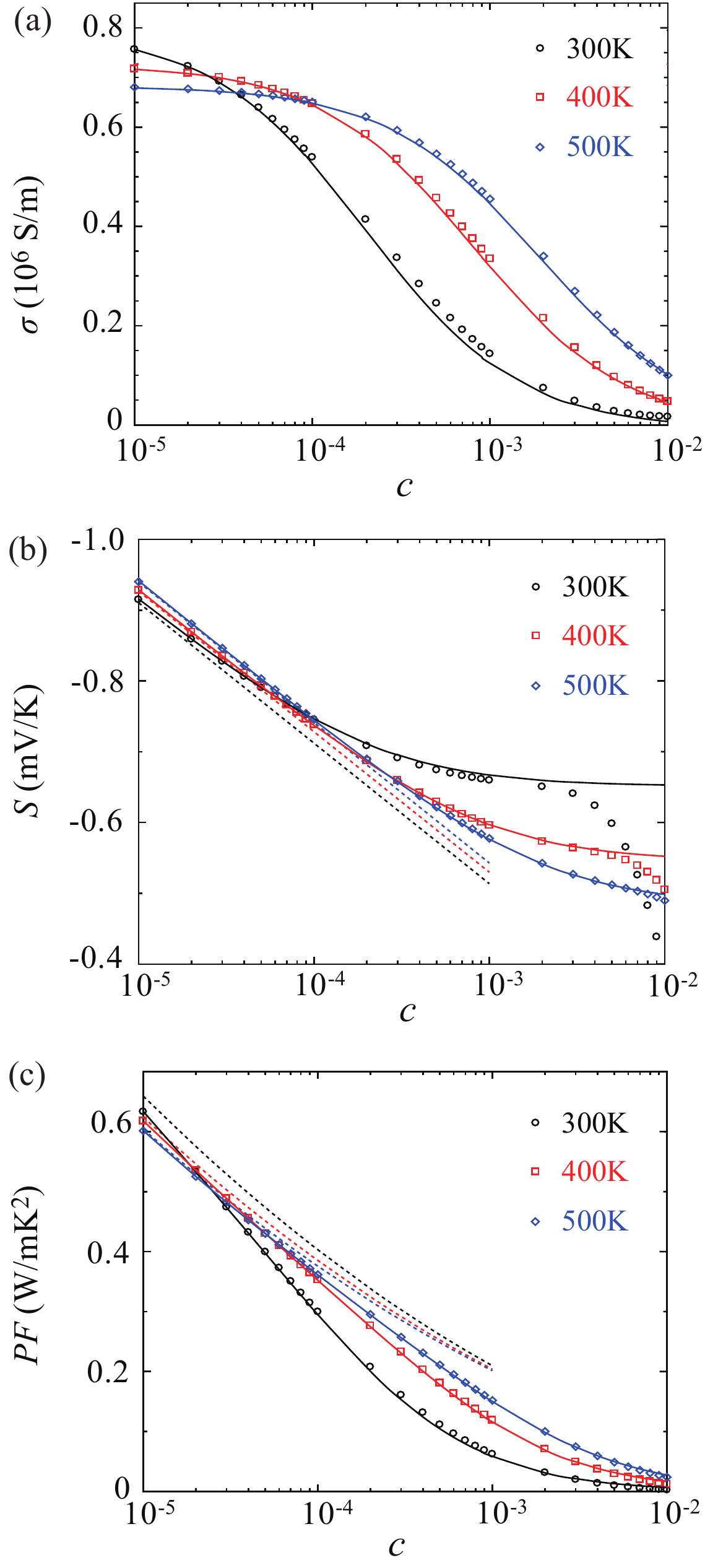}
  \end{center}
\caption{(Color online) The $c$-dependence of (a) the electrical conductivity $\sigma(=L_{11})$, (b) the Seebeck coefficient $S$, and 
(c) the power factor $PF$ of N-substituted (10,0) CNTs for which $c=0.01$, $0.03$, $0.05$, and $0.1$ at $T=300$~K ($\circ$), 
$400$~K ($\Box$), and $500$~K ($\diamond$). Note that the vertical axis of Fig.~\ref{fig:08}(b) is negative. 
The symbols ($\circ$, $\Box$, $\diamond$) and the solid curves indicate the results calculated based on the self-consistent $t$-matrix approximation
and on the Boltzmann formula in Eqs.~(\ref{eq:L11_2}) and (\ref{eq:L12_2}), respectively.
The dashed curves denote the asymptotic behaviors of $\sigma$, $S$ and $PF$ in the exhaustion region, respectively.}
\label{fig:08}
\end{figure}

Symbols ($\circ$, $\Box$, $\diamond$) in Fig.~\ref{fig:08}(a) present the $c$-dependence of electrical conductivity $\sigma(=L_{11})$ 
at $T=300$~K ($\circ$), 400~K ($\Box$), 500~K ($\diamond$) in the low-$c$ region where $c\le 0.01$. As $c$ decreases, $\sigma$ 
increases for all temperatures and eventually converges to a constant value in the limit of $c\to{0}$. This constant value of $\sigma$ 
decreases as $T$ increases. The reason of this low-$c$ behavior of $\sigma$ is explained in the next subsection (\S~\ref{sec:3.5}).

Symbols ($\circ$, $\Box$, $\diamond$) in Fig.~\ref{fig:08}(b) plot the $c$-dependence of the Seebeck coefficient $S$ 
at $T=300$~K ($\circ$), 400~K ($\Box$), 500~K ($\diamond$) in the low-$c$ region. As $c$ decreases, the absolute value of 
$S$ increases for all temperatures and eventually increases in proportion to $|\ln c|$, as indicated by the dotted lines in 
Fig.~\ref{fig:08}(b). Symbols ($\circ$, $\Box$, $\diamond$) in Fig.~\ref{fig:08}(c) show the $c$-dependence 
of the power factor $PF$ at $T=300$~K ($\circ$), 400~K ($\Box$), 500~K ($\diamond$) in the low-$c$ region. 
As $c$ decreases, $PF$ becomes large for a fixed temperature and eventually increases in proportion to $|\ln c|^2$, 
as seen by the dotted lines in Fig.~\ref{fig:08}(c). The following subsection (\S~\ref{sec:3.5}) elucidates the origin of the low-$c$ 
behavior of $S$ and $PF$.

An important aspect of the above results is that both $\sigma$ and $S$ increase with decreasing $c$ for fixed temperature, resulting 
in extremely high $PF$ values. This is in contrast to the conventional tradeoff relation between $\sigma$ and $S$ in terms of carrier 
concentration~\cite{rf:Mahan1998}. 

\subsection{Thermoelectric responses in the exhaustion region~\label{sec:3.5}}
At this point, we address the $c$-dependence of $L_{11}$ and $L_{12}$, as well as those of $S$ and $PF$, of N-substituted CNTs 
in the exhaustion region in which all N atoms (acting as donors) are ionized at high $T$. In the low-$c$ and high-$T$ regions, 
the transport phenomena are dominated by thermally-excited electrons in the conduction bands, whereas the contribution of impurity-band
electrons to $L_{11}$ and $L_{12}$ is negligible. In addition, the life time $\tau$ of conduction-band electrons is expected to be prolonged 
because of weak scattering from impurities. Therefore, the retarded self-energy $\Sigma^{\rm R}(E)$ for $E>0$ can be well described by
\begin{eqnarray}
\Sigma^{\rm R}(E)=-i\frac{\hbar}{2\tau(E)},
\label{eq:G^r_tau}
\end{eqnarray}
and $\alpha(E)$ in Eq.~(\ref{eq:alpha_E}) can be written as the Boltzmann expression
\begin{eqnarray}
\alpha(E)&=&\frac{4e^2\hbar}{\pi V}\sum_{k}v_{k}^2\left(\frac{\frac{\hbar}{2\tau(E)}}{{(E-\epsilon_{k})^2+\left(\frac{\hbar}{2\tau(E)}\right)^2}}\right)^2
\label{eq:alpha_E_tau0}\\
&=&\frac{4e^2N_{\rm unit}}{V}v^2(E)\tau(E)\rho_0(E),
\label{eq:alpha_E_tau}
\end{eqnarray}
where the factor 4 in Eqs.~(\ref{eq:alpha_E_tau0}) and (\ref{eq:alpha_E_tau}) reflects the spin and orbital degeneracies, and 
$\rho_0(E)$ is the DOS of a clean system ($c=0$) per the unit cell for each spin and each orbital, defined by
\begin{eqnarray}
\rho_0(E)=\frac{1}{N_{\rm unit}}\sum_{k}\delta(E-\epsilon_{k}).
\label{eq:rho_one}
\end{eqnarray}

Substituting Eq.~(\ref{eq:alpha_E_tau}) into Eqs.~(\ref{eq:L11}) and (\ref{eq:L12}), $L_{11}$ and $L_{12}$ reduce to the familiar 
Boltzmann formula within the relaxation-time approximation, as below.
\begin{eqnarray}
L_{11}\approx\frac{4e^2}{Aa_z}\int_{0}^{\infty}\!\!\!dE\left(-\frac{\partial f(E)}{\partial E}\right)v^2(E)\tau(E)\rho_0(E)
\label{eq:L11_2}
\end{eqnarray}
and 
\begin{eqnarray}
L_{12}\approx-\frac{4e}{Aa_z}\int_{0}^{\infty}\!\!\!dE\left(-\frac{\partial f(E)}{\partial E}\right)(E-\mu)v^2(E)\tau(E)\rho_0(E).
\label{eq:L12_2}
\end{eqnarray}
Here, it should be noted that the lower limit of the integrals in Eqs.~(\ref{eq:L11_2}) and (\ref{eq:L12_2}) is taken to be zero 
because the transport phenomena are expected to be dominated by the conduction-band electrons with energy $E\ge 0$ 
in the low-$c$ and high-$T$ regions as mentioned above. The validity of this approximation is confirmed by comparing
the results (solid curves in Fig.~\ref{fig:08}) obtained from Eqs.~(\ref{eq:L11_2}) and (\ref{eq:L12_2}) against the results 
(symbols in Fig.~\ref{fig:08}) calculated based on the self-consistent $t$-matrix approximations
including the contribution of both conduction- and impurity-band electrons to $L_{11}$ and $L_{12}$.
The calculation method of the solid curves in Fig.~\ref{fig:08} will be explained in detail below.

\begin{figure}[t]
  \begin{center}
  \includegraphics[keepaspectratio=true,width=70mm]{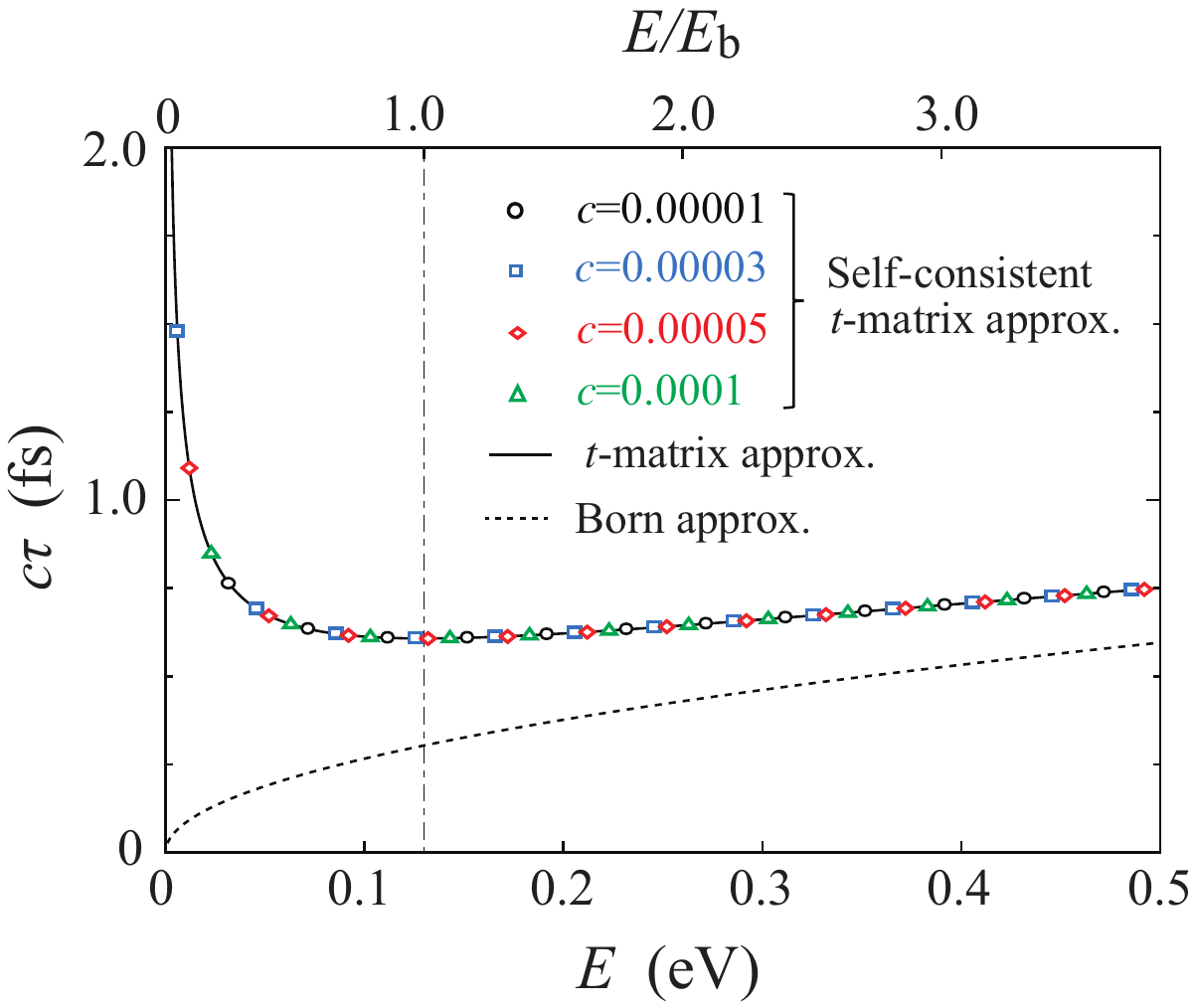}
  \end{center}
\caption{(Color online) The energy dependence of the life-times $\tau(E)$ of N-substituted (10,0) CNTs with low concentrations 
of N atoms ($0.00001$ ($\circ$), $0.00003$ ($\Box$), $0.00005$ ($\diamond$) and $c=0.0001$ ($\triangle$)), calculated 
within the self-consistent $t$-matrix approximation. The solid curve is the relaxation time in Eq.~(\ref{eq:tau_NSCT}) calculated
based on the $t$-matrix approximation. The dotted curve represents the relaxation time calculated using Fermi's golden rule based on 
the Born approximation. $E_{\rm b}$ is the absolute value of the binding energy of a single nitrogen atom in a z-CNT.}
\label{fig:09}
\end{figure}

Within the effective-mass approximation, the $v(E)$ and $\rho_0(E)$ values of a z-CNT for $E>0$ are respectively given by
\begin{eqnarray}
v(E)=\sqrt{\frac{2E}{m^*}}
\label{eq:vE_zCNT}
\end{eqnarray}
and
\begin{eqnarray}
\rho_0(E)=\frac{1}{\pi V_0}\sqrt{\frac{E_{\rm b}}{E}}.
\label{eq:rho_0E_zCNT}
\end{eqnarray}
In addition, the life time $\tau(E)$ for $E>0$ can be calculated as
\begin{eqnarray}
\tau^{-1}(E)&=&c\frac{2\pi}{\hbar}\left| \mathcal{T}(E)\right|^2\rho_0(E)
\label{eq:tau_NSCT0}\\
&=&\frac{2c}{\hbar}\frac{V_0}{\sqrt{E/E_{\rm b}}+\sqrt{E_{\rm b}/E}}
\label{eq:tau_NSCT}
\end{eqnarray}
within the $t$-matrix approximation (self-consistency is not necessary here because $c\to 0$). In Eq.~(\ref{eq:tau_NSCT}), 
we used $\mathcal{T}(E)=V_0/(1+i\sqrt{E_{\rm b}/E})$ as the $t$-matrix. It is noted that Eq.~(\ref{eq:tau_NSCT}) becomes 
$\tau(E)\approx \frac{\hbar}{2cV_0}\sqrt{E/E_{\rm b}}$ for $E\gg E_{\rm b}$, which is equivalent to the result of 
Fermi's golden rule $\tau^{-1}(E)=c\frac{2\pi}{\hbar}V_0^2\rho_0(E)$ based on the lowest-order Born approximation.
As shown in Fig.~\ref{fig:09}, the $t$-matrix approximation (the solid curve) is in excellent agreement with the $\tau$ values
calculated using the {\it self-consistent} $t$-matrix approximations, whereas the Born approximation (the dotted curve) 
deviates substantially from the numerical data. The breakdown of the Born approximation near the band edge is a particular 
feature of 1D semiconductors and originates from the van Hove singularity.

In addition, in the case of $E-\mu\gg k_{\rm B}T$, the Fermi-Dirac distribution function $f(E)$ can be approximately replaced by
the Boltzmann distribution function,
\begin{eqnarray}
f_{\rm B}(E-\mu)\approx e^{-\beta(E-\mu)}.
\label{eq:rho_0_zCNT}
\end{eqnarray}
Using Eqs.~(\ref{eq:vE_zCNT}), (\ref{eq:rho_0E_zCNT}), (\ref{eq:tau_NSCT}) and (\ref{eq:rho_0_zCNT}), the common term
$\left(-\frac{\partial f(E)}{\partial E}\right)v^2(E)\tau(E)\rho_0(E)$ in both Eqs.~(\ref{eq:L11_2}) and (\ref{eq:L12_2}) can be rewritten as
\begin{eqnarray}
\left(-\frac{\partial f(E)}{\partial E}\right)v^2(E)\tau(E)\rho_0(E)
=\frac{a_z^2\beta}{2c\pi\hbar}e^{-\beta(E-\mu)}\left(1+\frac{E}{E_{\rm b}}\right)
\end{eqnarray}
and $L_{11}$ and $L_{12}$ can be analytically calculated as
\begin{eqnarray}
L_{11}=\frac{2e^2a_z}{\pi\hbar A c}e^{\beta\mu}\left(1+\frac{1}{\beta E_{\rm b}}\right)
\label{eq:L11_mu}
\end{eqnarray}
and 
\begin{eqnarray}
L_{12}=-\frac{2ea_z}{\pi\hbar A c}\frac{e^{\beta\mu}}{\beta}\left(1+\frac{2}{\beta E_{\rm b}}\right)+\frac{\mu}{e}L_{11}.
\label{eq:L12_mu}
\end{eqnarray}
In addition, the temperature dependence of $\mu$ can be determined by Eq.~(\ref{eq:DOS2}). Because the persistence-type DOS for 
small $c$ values consists of both an impurity-band and conduction-band DOS,
\begin{eqnarray}
\rho(E)=\rho_{\rm imp}(E)+\rho_0(E),
\label{eq:rho_model}
\end{eqnarray}
the electron density $n$($=c/4$) for each spin and each orbital is calculated from Eq.~(\ref{eq:DOS3}) as
\begin{eqnarray}
\frac{c}{4}=ce^{\beta(E_{\rm b}+\mu)}+\frac{1}{2}\frac{e^{\beta\mu}}{\sqrt{\pi \beta t}},
\label{eq:n_exact}
\end{eqnarray}
where we note that $\rho_{\rm imp}(E)\approx c\delta(E+E_{\rm b})$ is a good approximation for $c\ll 1$, as shown in Fig.~\ref{fig:04}.
As seen from Eq.~(\ref{eq:n_exact}), the chemical potential $\mu$ obeys
\begin{eqnarray}
e^{\beta\mu}=\frac{c}{4}\left(ce^{\beta E_{\rm b}}+\frac{1}{2}\frac{1}{\sqrt{\pi \beta t}}\right)^{-1}.
\label{eq:e^betamu}
\end{eqnarray}
Substituting Eq.~(\ref{eq:e^betamu}) into Eqs.~(\ref{eq:L11_mu}) and (\ref{eq:L12_mu}), we obtain $L_{11}$ and $L_{12}$
as well as $S$ and $PF$ within the framework of Boltzmann transport theory, and the calculated $L_{11}(=\sigma)$, $S$, and
$PF$ are described by the solid curves in Figs.~\ref{fig:08}(a)-\ref{fig:08}(c), respectively. We can see in Figs.~\ref{fig:08}(a)-\ref{fig:08}(c) 
that the solid curves are in excellent agreement with the numerical data calculated based on the self-consistent $t$-matrix approximation 
in the low-$c$ region. The deviation of the solid curves from the symbols in the relatively large-$c$ region in Figs.~\ref{fig:08}(a)-\ref{fig:08}(c)
comes from the contribution of impurity-band electrons to $L_{11}$ and $L_{12}$

We now discuss the asymptotic behaviors of $\sigma$, $S$ and $PF$ in the limit of exhaustion region, respectively.
In the exhaustion region, the electron density in the impurity band (the first term on the right hand side in Eq.~(\ref{eq:n_exact})) 
is negligibly small in comparison with that in the conduction band (the second term on the right hand side in Eq.~(\ref{eq:n_exact})), 
which implies
\begin{eqnarray}
c\ll\frac{e^{-\beta{E_{\rm b}}}}{2\sqrt{\pi\beta t}}.
\label{eq:c_condition}
\end{eqnarray}
This is the condition that $c$ and $T$ must satisfy in the exhaustion region. Thus, in the exhaustion region, Eq.~(\ref{eq:n_exact}) 
leads to
\begin{eqnarray}
e^{\beta\mu}=\frac{c}{2}\sqrt{\pi\beta t}
\label{eq:e_exhaustion}
\end{eqnarray}
and, eventually, the temperature dependence of $\mu$ is given by
\begin{eqnarray}
\mu=\frac{1}{\beta}\ln\left(\frac{c}{2}\sqrt{\pi\beta t}\right).
\label{eq:beta_mu_exhaustion}
\end{eqnarray}
Note that Eq.~(\ref{eq:c_condition}) and Eq.~(\ref{eq:beta_mu_exhaustion}) indicate that $\mu$ is 
much lower than the bound-state energy $E=-E_{\rm b}$ in the exhaustion region ({\it i.e.}, $\mu\ll -E_{\rm b}$).

Substituting Eq.~(\ref{eq:e_exhaustion}) into Eq.~(\ref{eq:L11_mu}), the conductivity $L_{11}$ becomes
\begin{eqnarray}
L_{11}=\frac{e^2a_z}{\hbar A}
\sqrt{\frac{t}{\pi k_{\rm B}T}}
\left(1+\frac{k_{\rm B}T}{E_{\rm b}}\right).
\label{eq:L11_exhaustion}
\end{eqnarray}
It should be noted that $L_{11}(=\sigma)$ is independent of $c$ in the exhaustion region. This is the reason that 
$\sigma$ converges to a constant value in the limit of $c\to{0}$ as shown in Fig.~\ref{fig:08}(a)). 
This results from the cancellation between the life-time $\tau\propto 1/c$ and the conduction carrier density ($\propto c$), 
as seen in the derivation of Eq.~(\ref{eq:L11_exhaustion}). This is a unique characteristic of 1D semiconductors arising from 
the van Hove singularity of the DOS. In the case of N-substituted (10,0) CNTs, $E_{\rm b}$ is much higher than thermal energy
corresponding to the maximum temperature of 500K considered in the present study, {\it i.e.}, $k_{\rm B}T/E_{\rm b}\ll 1$. In this case, 
the conductivity $\sigma(=L_{11})$ in Eq.~(\ref{eq:L11_exhaustion}) decreases as $T$ is raised.

\begin{figure}[t]
  \begin{center}
  \includegraphics[keepaspectratio=true,width=70mm]{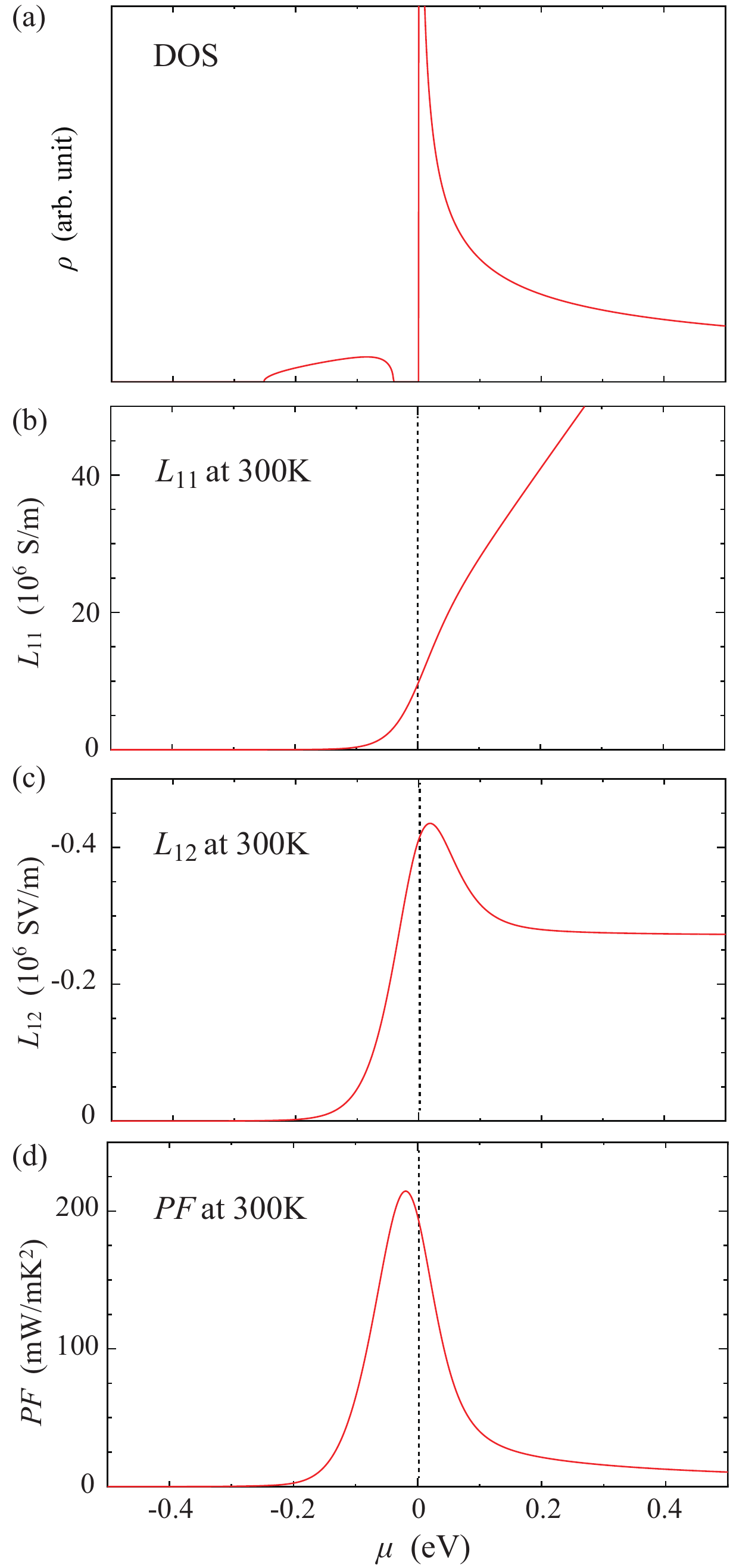}
  \end{center}
\caption{(Color online) The $\mu$ dependence of the (a) DOS, (b) $L_{11}$ at 300~K, (c) $L_{12}$ at $300$~K, and (d) $PF$ 
at $300$~K of N-substituted (10,0) CNTs for which $c=0.01$. The origin of the horizontal axis ($\mu=0$~eV) corresponds to 
the conduction-band edge and is represented by the dashed lines in Figs.~\ref{fig:10}(b)-\ref{fig:10}(d).}
\label{fig:10}
\end{figure}

Similarly, $L_{12}$ can be obtained as 
\begin{eqnarray}
L_{12}=-\frac{ea_z}{\hbar A}\sqrt{\frac{t k_{\rm B}T}{\pi}}\left(1+2\frac{k_{\rm B}T}{E_{\rm b}}\right)+\frac{\mu}{e}L_{11},
\label{eq:L12_exhaustion}
\end{eqnarray}
where $\mu$ and $L_{11}$ are given by Eqs.~(\ref{eq:e_exhaustion}) and (\ref{eq:L11_exhaustion}), respectively. 
In the present case, which satisfies $k_{\rm B}T/E_{\rm b}\ll 1$, the second term in parentheses on the right hand side in Eq.~(\ref{eq:L12_exhaustion})
is negligible. Substituting Eqs.~(\ref{eq:L11_exhaustion}) and (\ref{eq:L12_exhaustion}) into Eq.~(\ref{eq:S}), the Seebeck coefficient 
$S$ is expressed as
\begin{eqnarray}
S=-\frac{k_{\rm B}}{e}\left\{
\frac{1+2k_{\rm B}T/E_{\rm b}}{1+k_{\rm B}T/E_{\rm b}}+\frac{1}{2}\ln\left(\frac{k_{\rm B} T}{t}\right)-\ln\left(\frac{\sqrt{\pi}}{2}c\right)
\right\}. 
\label{eq:S_exhaustion}
\end{eqnarray}
For $k_{\rm B}T/E_{\rm b}\ll 1$, the first term in the curly brackets on the right hand side in Eq.~(\ref{eq:S_exhaustion})
can be regarded as $\sim$1 and eventually $S\propto -\ln(k_{\rm B}T/t)$ (see also the dashed lines in Fig.~\ref{fig:08}(b)). 
For a fixed $T$, the Seebeck coefficient $S$ in the exhaustion region increases logarithmically with $c$, such that $S\propto\ln c$ 
(see also the dashed lines in Fig.~\ref{fig:08}(b)). From the results of $\sigma=$const (with respect to $c$) and $S\propto\ln{c}$, 
we can see that the power factor $PF$ behaves as $PF\propto(\ln{c})^2$ in the exhaustion region (see also the dashed lines in 
Fig.~\ref{fig:08}(c)).

\subsection{Chemical potential dependence of $L_{11}$, $L_{12}$ and $PF$~\label{sec:3.6}}
Lastly, we discuss the $\mu$-dependence of the thermoelectric responses of N-substituted z-CNTs. In experimental work, 
the chemical potential $\mu$ of CNTs can be adjusted by chemical adsorption on a CNT surface~\cite{rf:nakai,rf:nonoguchi1,rf:fujigaya2}, 
encapsulation of molecules inside a CNT~\cite{rf:fujigaya1}, or carrier injection into CNTs by applying a gate voltage using a field-effect 
transistor (FET) setup~\cite{rf:yanagi}. Thus, $L_{11}$ and $L_{12}$ as functions of $\mu$ are of interest.

Figure~\ref{fig:10}(a) plots the $\mu$ dependence of the DOS for N-substituted (10,0) CNTs for which $c=0.01$, which is the same as 
the DOS in Fig.~\ref{fig:03}(a). Figures~\ref{fig:10}(b)-\ref{fig:10}(d) show similar plots for the $L_{11}$, $L_{12}$ and $PF$ values of 
N-substituted (10,0) CNTs for which $c=0.01$, respectively. The conduction-band edge is denoted by the vertical dashed lines in 
Fig.~\ref{fig:10}(b)-\ref{fig:10}(d). As seen in Fig.~\ref{fig:10}(b), $L_{11}$ starts to increase from just below the conduction-band edge, 
and is proportional to $\mu$ when $\mu$ lies far above the conduction-band edge. In contrast, $L_{12}$ exhibits a peak in the vicinity of 
the conduction-band edge and is constant when $\mu$ is located far above the conduction-band edge. Substituting these $L_{11}$ and 
$L_{12}$ data into Eq.~(\ref{eq:PF}), the $\mu$ dependence of $PF$ can be obtained as shown in Fig.~\ref{fig:10}(d), in which $PF$ exhibits 
a maximum value near the conduction-band edge. A key aspect of these results is that ``band-edge engineering" is essential for 
the enhancement of the thermoelectric performances of such materials.

\section{Summary and Conclusions~\label{sec:4}}
The thermoelectric properties of N-substituted semiconducting CNTs were theoretically investigated using exact expressions 
for thermoelectric response functions. We found that the power factor ($PF$) values of N-substituted CNTs increase with decreases in 
the N concentration and that extremely high values are eventually obtained in the exhaustion region. These high values result from both 
monotonic increase in the Seebeck coefficient $S$ and saturation in the electrical conductivity $\sigma$ with decreasing N concentration 
(see Fig.~\ref{fig:08}). This is in contrast to the typical inverse relation between $S$ and $\sigma$ in conventional bulk semiconductors 
and is a unique feature of 1D semiconductors originating from the van Hove singularity of the DOS. Thus, we conclude that 
1D semiconductors such as semiconducting CNTs are promising candidates for high-performance thermoelectric materials as proposed
by Hicks and Dresselhaus in 1993~\cite{rf:hicks}.

Furthermore, inspired by recent experiments regarding carrier doping effects on the thermoelectric properties of CNTs~\cite{rf:nakai,rf:yanagi,rf:nonoguchi1,rf:nonoguchi2,rf:nonoguchi3,rf:fujigaya1,rf:fujigaya2}, we studied the chemical-potential 
dependence of the thermoelectric responses of z-CNTs and found that the $PF$ values exhibit a maximum in the vicinity of the band edge. 
A similar $PF$ peak around the band edge has been observed in recent experimental work with thin sheets made from aggregates of 
the semiconducting CNTs (so-called buckypapers)~\cite{rf:yanagi}. Similar experimental studies for an individual CNT are desired.
A key aspect of these results is that ``band-edge engineering" is essential for 
the enhancement of the thermoelectric performances of materials.

\section*{Acknowledgements}
\begin{acknowledgment}
The authors are grateful to Mildred S. Dresselhaus for her stimulating our interest on thermoelectric effects of one-dimensional 
materials. H.F. also thanks her for her constant encouragement since early 1970s. The authors also thank Masao Ogata, 
Hiroyasu Matsuura, Hideaki Maehashi and Kenji Sasaoka for valuable discussions during this work. This work 
was supported, in part, by a JSPS KAKENHI grant (no. 15H03523).
\end{acknowledgment}

\appendix

\section{Impurity-concentration dependence of $\alpha(\epsilon_{\rm F})$, $\alpha'(\epsilon_{\rm F})$ and $\alpha''(\epsilon_{\rm F})$}
\begin{figure}[h]
  \begin{center}
  \includegraphics[keepaspectratio=true,width=70mm]{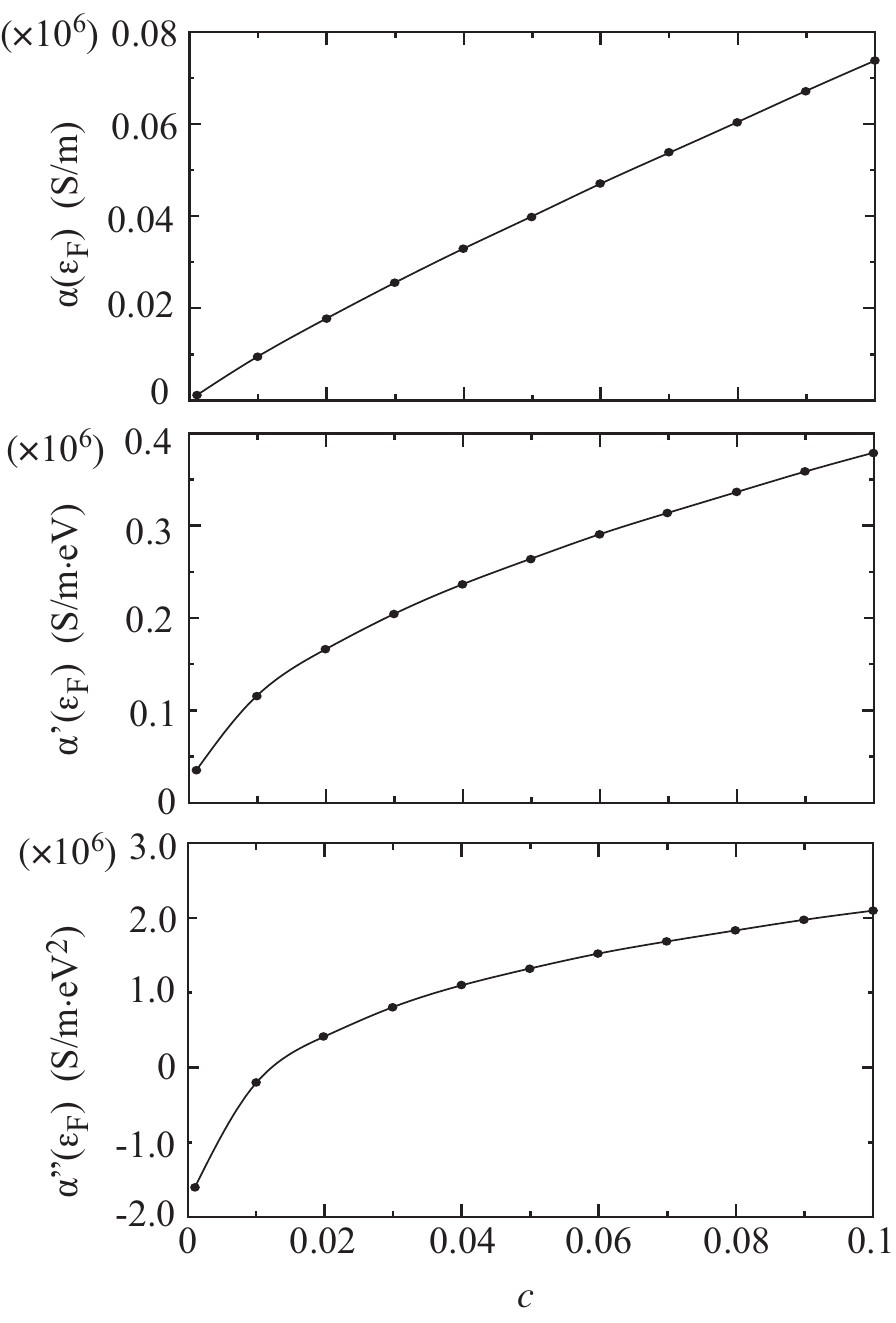}
  \end{center}
\caption{
$c$-dependence of (a) $\alpha(\epsilon_{\rm F})$, (b) $\alpha'(\epsilon_{\rm F})$ and (c) $\alpha''(\epsilon_{\rm F})$
for N-substituted (10,0) CNTs.
}
\label{fig:A1}
\end{figure}

As indicated in \S~\ref{sec:3.2} and \S~\ref{sec:3.3}, the low-temperature behavior of $L_{11}$ and $L_{12}$ as well as $S$ and $PF$ are
determined by $\alpha(\epsilon_{\rm F})$, $\alpha'(\epsilon_{\rm F})$ and $\alpha''(\epsilon_{\rm F})$. Figure~\ref{fig:A1} represents
$c$-dependence of (a) $\alpha(\epsilon_{\rm F})$, (b) $\alpha'(\epsilon_{\rm F})$ and (c) $\alpha''(\epsilon_{\rm F})$. As shown in
Fig.~\ref{fig:A1}, $\alpha(\epsilon_{\rm F})$ and $\alpha'(\epsilon_{\rm F})$ are positive and monotonically increase with respect to $c$ 
which are reflected in the low-temperature limit of $L_{11}$ and $T^2$ coefficients of $L_{12}$. On the other hand $\alpha''(\epsilon_{\rm F})$ 
is negative for small $c$, changes signs and then monotonically increases as $c$ increases, which explain the slight decrease of $L_{11}$ 
as temperatures is raised at $c=0.01$.

\end{document}